\documentclass[]{aa}

\usepackage{graphicx,url,twoopt,natbib,siunitx,lipsum,footnote,threeparttable}
\usepackage[varg]{txfonts}           
\usepackage{hyperref}
\usepackage[labelfont={rm,bf,up}]{caption}
\usepackage{subfig, placeins}
\hypersetup{colorlinks=true,citecolor=blue,urlcolor=blue} 

\usepackage[dvipsnames]{xcolor}
\bibpunct{(}{)}{;}{a}{}{,}   

\newcommand{\objectname}[0]{{\object{9~Sgr}}}
\newcommand{\Msun}[0]{{M_{\odot}}}
\newcommand{\Lsun}[0]{{L_{\odot}}}
\newcommand{\Rsun}[0]{{R_{\odot}}}
\newcommand{\kms}[0]{\si{\km\per\second}}
\newcommand{\teff}[0]{T_{\rm eff}}
\newcommand{\halpha}[0]{\element{H}$\alpha$}
\newcommand{\hgamma}[0]{\element{H}$\gamma$}
\newcommand{\hdelta}[0]{\element{H}$\delta$}
\newcommand{\C}[1]{[\element{C#1}/\element{H}] + 12}
\newcommand{\N}[1]{[\element{N#1}/\element{H}] + 12}
\renewcommand{\ll}[0]{\lambda\lambda}
\renewcommand{\O}[1]{[\element{O#1}/\element{H}] + 12}

\begin{document}

\title{Resolving the dynamical mass tension of the massive binary \objectname{}\thanks{Based on observations collected at the European Southern Observatory under program IDs 083.D-0066(A), 085.C-0389(B), 386.D-0198(A), 086.D-0586(B), 091.D-0334(A), 092.D-0590(B), 093.D-0673(A), 093.D-0673(C), 60.A-9209(A), 093.D-0039(A), 596.D-0495(D), 596.D-0495(J) \& 60.A-9158(A), and with the Mercator telescope, operated on the island of La Palma by the Flemish Community, at the Spanish Observatorio del Roque de los Muchachos of the Instituto de Astrofísica de Canarias.}}

\author{
M. Fabry\inst{\ref{ivs}} \and 
C. Hawcroft\inst{\ref{ivs}} \and 
A. J. Frost\inst{\ref{ivs}} \and 
L. Mahy\inst{\ref{rob},\ref{ivs}} \and 
P. Marchant\inst{\ref{ivs}} \and
J-B. Le Bouquin\inst{\ref{ipag}} \and 
H. Sana\inst{\ref{ivs}} 
}
\institute{
Institute of Astronomy (IvS), KU Leuven, Celestijnenlaan 200D, 3001 Leuven, Belgium\\ \email{matthias.fabry@kuleuven.be} \label{ivs} 
\and
Royal Observatory of Belgium (ROB), Avenue Circulaire 3, 1180 Brussels, Belgium \label{rob}
\and
Institute of Planetology and Astrophysics (IPAG), Grenoble University, Rue de la Piscine 414, 38400 St-Martin d’Hères, France \label{ipag}
}
\date{received date; accepted date}


\abstract
{
Direct dynamical mass measurements of stars with masses above $30\,\Msun{}$ are rare. This is the result of the low yield of the upper initial mass function and the limited number of such systems in eclipsing binaries. Long-period, double-lined spectroscopic binaries that are also resolved astrometrically offer an alternative to eclipsing binaries for obtaining absolute masses of stellar objects. \objectname{} (HD 164794) is one such long-period, high-mass binary. Unfortunately, a large amount of tension exists between its total dynamical mass inferred spectroscopically from radial velocity measurements and that from astrometric data.
}
{
Our goal is to resolve the mass tension of \objectname{} that exists in literature, to characterize the fundamental parameters and surface abundances of both stars, and to determine the evolutionary status of the binary system, henceforth providing a reference calibration point to confront evolutionary models at high masses.
}
{
We obtained the astrometric orbit from existing and new multi-epoch VLTI/PIONIER and VLTI/GRAVITY interferometric measurements. Using archival and new spectroscopy, we performed a grid-based spectral disentangling search to constrain the semi-amplitudes of the radial velocity curves. We computed atmospheric parameters and surface abundances by adjusting  \textsc{fastwind} atmosphere models and we compared our results with evolutionary tracks computed with the Bonn Evolutionary Code (BEC).
}
{
Grid spectral disentangling of \objectname{} supports the presence of a $53\,\Msun{}$ primary and a $39\,\Msun{}$ secondary, which is in excellent agreement with their observed spectral types. In combination with the size of the apparent orbit, this puts \objectname{} at a distance of $1.31\pm0.06\,\si{kpc}$. Our best-fit models reveal a large mass discrepancy between the dynamical and spectroscopic masses, which we attribute to artifacts from repeated spectral normalization before and after the disentangling process. Comparison with BEC evolutionary tracks shows the components of \objectname{} are most likely coeval with an age of roughly $1\,\si{\mega yr}$.
}
{
Our analysis clears up the contradiction between mass and orbital inclination estimates reported in previous studies. We detect the presence of significant CNO-processed material at the surface of the primary, suggesting enhanced internal mixing compared to currently implemented in the BEC models. The present measurements provide a high-quality high-mass anchor to validate stellar evolution models and to test the efficiency of internal mixing processes.
}


\maketitle

\section{Introduction} \label{sec:intr}
Massive stars drive the chemical enrichment of heavy elements and inject large amounts of kinetic energy into their neighborhoods through their strong, line-driven winds and final explosion as supernovae and gamma-ray bursts.
Obtaining accurate mass measurements of stars in the upper part of the Hertzsprung-Russell diagram (HRD) has been a challenge, however, as evolutionary models of high-mass stars are riddled with physical uncertainties.
Furthermore, spectroscopic masses, obtained through atmospheric model fitting, are intrinsically inaccurate.
Therefore, direct mass measurements that are independent of atmosphere or evolutionary models offer valuable constraints to gauge the quality of the models.\par

Binary stars are the prime laboratories to obtain accurate, model independent masses through Kepler's laws, which provide so-called dynamical masses.
Unfortunately, no single observational technique can fully characterize the orbit and dynamical masses of the two components.
Either a double-lined spectroscopic binary (SB2) has to be eclipsing, \emph{or} astrometric data, either absolute or relative, must be available. In both cases, multi-epoch observations are required.\par

Traditionally, eclipsing SB2s are considered to provide the best constraints on the orbital parameters.
Yet, they are rare and uncertainties about the effects of tidal deformation, mutual illumination and/or binary interaction may pollute the obtained results, making it challenging to confront these objects to single-star models.
This is particularly the case in the realm of massive stars.
In this context, astrometric binaries are a valuable alternative to eclipsing SB2s.
Recent advances in optical long-baseline interferometry have identified a number of such systems \citep[e.g.,][]{sanaThreedimensionalOrbitsTripleO2013, sanaSOUTHERNMASSIVESTARS2014, mayerThreebodySystemCircini2014, maizapellanizCloseEncounterMassive2017, mahyTripleSystemHD2018}, which offer new opportunities to obtain dynamical mass constraints of stars in the upper mass function.\par

\objectname{} (HD164794) is such a long-period astrometric SB2 system in the Lagoon Nebula.
It was first studied by \citet{abbottDetectionVariableNonthermal1984} in the context of its variable synchrotron emission, which is interpreted to be a result of wind-wind collisions of binaries \citep{pittardRadioXrayGray2006}, hinting that the then presumed single star \objectname{} was in fact a binary.
Subsequent studies by \citet{rauwMultiwavelengthInvestigationNonthermal2002, rauwXMMNewtonObservationLagoon2002} and \citet{nazeBinarySignatureNonthermal2008} confirmed the presence of elevated X-ray emission, a typical indication of colliding winds in O + O  binaries \citep{rauwPhaseresolvedXrayOptical2002, sanaPhaseresolvedXMMNewtonCampaign2004, sanaXMMNewtonViewYoung2006, rauwXrayEmissionInteracting2016}.
\citet{rauwSagittariiUncoveringOtype2012} first confirmed the long-period binary nature of \objectname{} through radial velocity (RV) measurements and classified its components as O3.5V((f\textsuperscript{+})) and  O5.5V((f)).
\citet{rauwTestingTheoryColliding2016} studied the periastron passage of \objectname{} and reported a maximum in the X-ray emission coming from shocked gas in the interaction zone of the stellar winds, as expected from a wind-wind collision zone in a wide binary where the shocked material cools adiabatically \citep{stevensCollidingWindsEarlytype1992}.\par

The system was resolved for the first time in 2009 using the Astronomical Multi-BEam combineR (AMBER) and in 2013 with  the Precision Integrated-Optics Near-infrared Imaging ExpeRiment (PIONIER) by \citet{sanaSOUTHERNMASSIVESTARS2014}.
\citet{lebouquinResolvedAstrometricOrbits2017} constrained the astrometric orbit using multi-epoch AMBER and PIONIER interferometric measurements, uncovering a discrepancy with the spectroscopic analyses of \cite{rauwSagittariiUncoveringOtype2012, rauwTestingTheoryColliding2016}.
While the interferometric measurements of \citet{lebouquinResolvedAstrometricOrbits2017} firmly excluded inclinations below 80 degrees, the RV curve semi-amplitudes of \citet{rauwSagittariiUncoveringOtype2012, rauwTestingTheoryColliding2016} resulted in an estimated inclination of about 50 degrees if the stars were to have masses representative of their spectral types.\par

Another long standing issue is the distance to \objectname{}, and whether or not it is a member of the young open cluster NGC~6530.
\citet{prisinzanoStarFormationRegion2005} and \citet{kharchenkoAstrophysicalParametersGalactic2005} measured distances to the cluster of around $1.25\,\si{kpc}$, while earlier measurements indicated a distance close to $1.8\,\si{kpc}$ \citep{vandenanckerMultiwavelengthStudyStar1997, sungUBVRIHaPhotometry2000} as a result of differences in the adopted reddening laws.
The combination of the astrometric and spectroscopic measurements provide a direct constraint on the distance, allowing us to confirm the current Gaia eDR3 measurement of $1.21\,\si{kpc}$.
Gaia does not suffer from systematics owing to an assumed reddening law, and thus provides a powerful constraint, but its measurement can still be impacted by the multiplicity of the system.\par

In this work, we aim to resolve the existing discrepant results that cast doubt on the current mass estimates, evolutionary status, and membership of \objectname{}.
To do so, we leverage the accuracy of relative astrometry and we perform a grid spectral disentangling analysis on spectroscopic data to fully constrain the orbit of \objectname{}.
We further use the atmospheric properties of the stars in the system to derive its evolutionary status and test stellar evolution models at high masses.
This paper is organized as follows.
Section~\ref{sec:obs} covers the observational data that were used.
We present and discuss the orbital analysis and spectral disentangling in Sect.~\ref{sec:orb}, the atmosphere modeling of both components in Sect.~\ref{sec:atm}, and the evolutionary status of \objectname{} is discussed in Sect.~\ref{sec:evol}.
Section~\ref{sec:conc} presents our conclusions and final remarks.\par
\section{Observations}\label{sec:obs}
We combine archival and new optical spectroscopy with near-infrared (NIR) interferometry of \objectname{}.
Most of these measurements were part of long-term monitoring programs.\par

\subsection{Optical spectra}\label{sec:optspec}
Archival data consist of 57 spectra from the High Efficiency and Resolution Mercator Echelle Spectrograph (HERMES, \citealt{raskinHERMESHighresolutionFibrefed2011}, used in \citealt{rauwTestingTheoryColliding2016}), 20 Fiber-fed Extended Range Optical Spectrograph (FEROS) spectra \citep{kauferCommissioningFEROSNew1999} and 49 Ultraviolet and Visual Echelle Spectrograph (UVES) spectra \citep{dekkerDesignConstructionPerformance2000} in the blue and red arms \citep[used in][]{rauwSagittariiUncoveringOtype2012}.
Obsrevations that are not previously analyzed are listed in Table \ref{tab:newspectra} and consist of five additional HERMES spectra and one spectrum from the High Accuracy Radial velocity Planet Searcher (HARPS) spectrograph \citep{mayorSettingNewStandards2003}.
The FEROS spectra cover the spectral domain between 3700 and $9000\,\si{\angstrom}$ and have a resolving power of $R\approx48\,000$.
The UVES spectra cover the wavelength range $\lambda\lambda = 3500-5000\,\si{\angstrom}$ with its blue arm and $\lambda\lambda = 5000-7000\,\si{\angstrom}$ with its red arm, and each have a resolving power of $R\approx40\,000$.
The HERMES and HARPS spectra both have $R\approx85\,000$ with a coverage of $\lambda\lambda = 3800-9000\,\si{\angstrom}$ and $\lambda\lambda = 3750-6900$, respectively.
The FEROS, UVES, and HARPS spectra were obtained through the ESO archive science portal and were pre-reduced with their respective pipelines.
The HERMES spectra are reduced using the HERMES Data Reduction Software (DRS) pipeline.
Finally, all spectra were normalized over their whole spectral domain by fitting a cubic spline function through selected continuum regions.\par

\subsection{Near-infrared Interferometry}\label{sec:interf}
We used the previously published AMBER and PIONIER dataset obtained from Jun 2009 to Aug 2016 \citep{lebouquinResolvedAstrometricOrbits2017}, along with two new PIONIER observations obtained in May and August 2017.
These new data show for the first time the system turning back on the apparent ellipse and provide an almost complete coverage of the nine-year orbit.
They were obtained with PIONIER \citep{lebouquinPIONIER4telescopeVisitor2011} at the Very Large Telescope Interferomer (VLTI) using the four 1.8 meter Auxiliary Telescopes (ATs) in configurations B2-K0-D0-I3 and A0-G1-J2-J3, offering a maximal projected baseline of 120 and 130 meter respectively.
The PIONIER data were reduced and analyzed, as described in \citet{lebouquinResolvedAstrometricOrbits2017}, using the \texttt{pndrs}\footnote{\url{http://www.jmmc.fr/pndrs}} package.
Each observation produces six visibilities and four closure phases, delivering relative astrometry with sub-milliarcsecond precision and an H band flux ratio of $f_{\rm H} = 0.62\pm0.02$.
The full journal of interferometric observations is given in Table~\ref{tab:newinter} along with the measured astrometric properties of the system. \par
Additionally, three VLTI/GRAVITY \citep{gravitycollaborationFirstLightGRAVITY2017} measurements were taken, of which two were obtained in June 2016 and the other in September 2016.
These observations were part of the science verification (SV) program and used the ATs in configurations A0-G1-D0-C1 and A0-G1-J2-K0.
As GRAVITY is a spectro-interferometer, it provides six visibilities and four closure phases for each wavelength bin in the $2.0-2.4\,\si{\micro\meter}$ K band with a spectral resolving power of $R=4000$.
These SV data were reduced with the standard GRAVITY pipeline (\citealt{lapeyrereGRAVITYDataReduction2014}, version 1.0.11) and fitted parametrically to a binary model with PMOIRED\footnote{\url{https://github.com/amerand/PMOIRED}}.
The uncertainties on the relative astrometry were estimated by adding a bootstrapped error and a systematic error of $0.1\si{mas}$ in quadrature.
In the bootstrapping procedure, data were drawn randomly to create new datasets and the final parameters and uncertainties were estimated as the average and standard deviation of all the fits that were performed.
Unfortunately, in this data the spectral signal was not of sufficient quality to infer RVs.


\begin{figure*}
	\resizebox{\hsize}{!}{\includegraphics{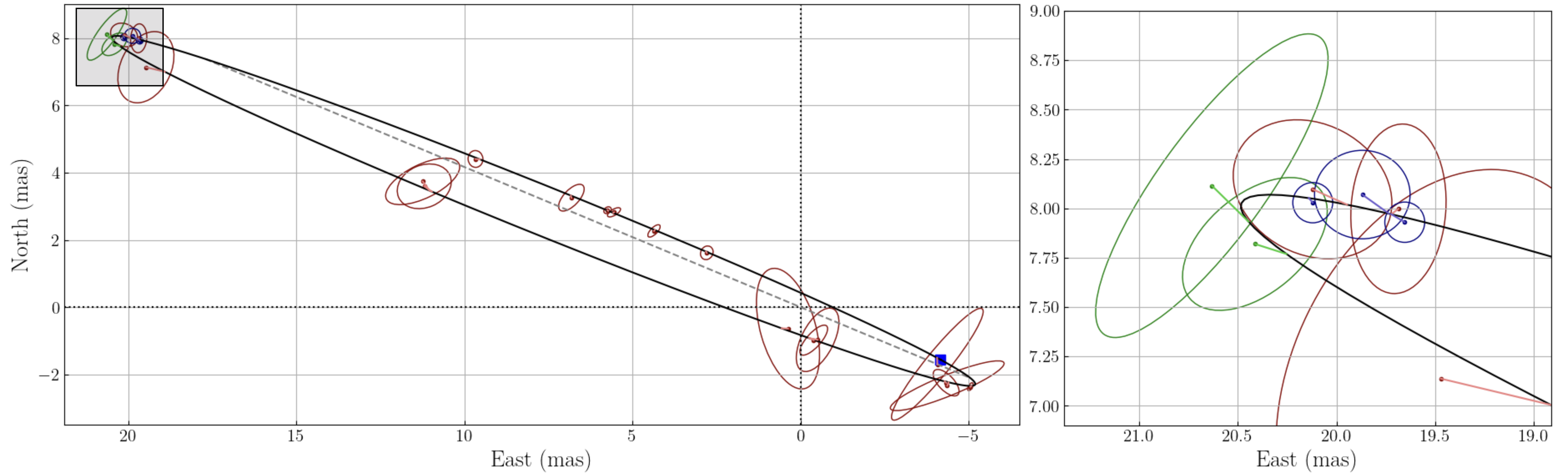}}
	\caption{Relative astrometric orbit of \objectname{}.
		Archival observations are denoted in red, the new PIONIER data in green, and the GRAVITY observations in blue, all with their respective error ellipses.
		The blue square indicates the periastron passage and the dashed line represents the line of nodes.
		The right panel shows the zoom-in of the shaded region.}
	\label{fig:asfig}
\end{figure*}

\begin{table}
	\caption{Relative astrometric solution of \objectname{}, where the usual notation of the orbital elements have been adopted (see Appendix~\ref{app:spinOS}). The reported errors correspond to the 1$\sigma$ confidence interval determined by the MCMC analysis.}
	\label{tab:orbsol}
	\centering
	\begin{tabular}{c | c c}
		\hline\hline
		Element [Unit] & Value & Error \\
		\hline
		$P\,[\si{\day}]$ & 3261 & 69\\
		$T_0\,[\si{MJD}]$ & 56547 & 12\\
		$\Omega\,[\si{deg}]$ & 67.3 & 0.4 \\
		$\omega\,[\si{deg}]$\tablefootmark{a} & 210.7 & 2.3\\
		$i\,[\si{deg}]$ & 86.5 & 0.5\\
		$e$ & 0.648 & 0.009 \\
		\hline
	\end{tabular}
	\tablefoot{
		\tablefoottext{a}{By construction, the periastron argument of the relative orbit $\omega$ is shifted by 180\degr{} compared to the argument of periastron $\omega_1$ that is traditionally adopted by fitting the RV curve of the primary star.}
	}
\end{table}

\section{Orbital analysis}\label{sec:orb}
\subsection{Astrometric orbit}\label{sec:orbit}
The interferometric observations allow us to constrain the orbital parameters of the relative orbit, including its orbital period ($P$) and eccentricity ($e$), by solving for the Thiele-Innes constants; see Eqs. \eqref{eq:thielea}--\eqref{eq:thieleg}.
We do so through a Levenberg-Marquardt least-squares minimization algorithm implemented in the newly developed Python package \textsc{spinOS}\footnote{\url{https://github.com/matthiasfabry/spinOS}} that is described in Appendix \ref{app:spinOS}.
Errors on each parameter are determined by running a Markov chain Monte Carlo (MCMC) analysis of a hundred thousand samples.
The resulting apparent orbit and orbital elements are shown in Fig. \ref{fig:asfig} and Table \ref{tab:orbsol}.
The scatterplot matrix of the MCMC analysis is deferred to Fig. \ref{fig:corner}.\par

As \citet{lebouquinResolvedAstrometricOrbits2017} find, the orbit is very well constrained with a high eccentricity $e=0.648\pm0.009$, a near edge-on inclination of the orbital plane with the plane of the sky of $i=86.5^\circ\pm0.5^\circ$ and an orbital period of $P=8.9\pm0.2$ years.
The scatterplot matrix of the MCMC confirms this solution is robust (Fig. \ref{fig:corner}).
Importantly, we corroborate the findings of \citet{lebouquinResolvedAstrometricOrbits2017} that inclinations of $85^\circ$ or lower are ruled out.\par

\begin{table}
	\caption{Spectral lines and their respective wavelength ranges that were disentangled to do the grid search described in section \ref{sec:disen}.}
	\label{tab:fd3gridlines}
	\centering
	\begin{tabular}{c c c c}
		\hline\hline
		Spectral line & $\lambda_{\rm min}[\si{\angstrom}]$ & $\lambda_{\rm max}[\si{\angstrom}$]\\
		\hline
		$\ion{He}{i}\, 4026$ & 4020.0 & 4031.0\\
		$\element{H}\delta$ & 4091.8 & 4112.5\\
		$\ion{He}{ii}\, 4200$ & 4193.0 & 4206.0 \\
		$\element{H}\gamma$ & 4327.8 & 4354.5 \\
		$\ion{He}{i}\, 4471$ & 4465.0 & 4477.0\\
		$\ion{He}{ii}\, 4541$ & 4532.5 & 4550.0 \\
		$\ion{He}{ii}\, 4686$ & 4677.0 & 4692.0 \\
		$\element{H}\beta$ & 4841.6 & 4878.0\\
		$\ion{He}{i}\, 4922$ & 4917.0 & 4927.0\\
		$\ion{He}{i}\, 5016$ & 5011.0 & 5026.0\\
		$\ion{He}{ii}\, 5411$ & 5399.2 & 5424.1 \\
		$\ion{He}{i}\, 5875$ & 5871.0 & 5879.5\\
		\hline
	\end{tabular}
\end{table}

\subsection{Spectral disentangling}\label{sec:disen}
The astrometric solution of Sect. \ref{sec:orbit} only provides the relative size of the orbit.
Determining the absolute size of the orbit independently of an assumed distance requires additional information about the system.
In the case of \objectname{}, this information is contained in the Doppler motion of its spectra, specifically, in the values of the semi-amplitudes of its RV curves ($K_1$ and $K_2$).
In the standard approach, we would fit the profiles of selected spectral lines, or determine the line barycenter, and measure the red- or blue shift as a function of the observation time to obtain an observational RV curve.\par

However, in long-period systems with broad spectral lines such as \objectname{}, the lines of the two components may never fully deblend, and a small amount of contamination by the other component may significantly impact the measured RVs and therefore the orbital solution and derived masses.
Spectral disentangling can help to improve the measured RVs in this respect because it is able to account for the cross-contamination of the spectral lines of the components.
Spectral disentangling however still faces challenges when the spectral lines never fully deblend or when the orbital solution is poorly defined.
The most advanced spectral disentangling algorithms can optimize the orbital solution at the same time as separating the spectra, but the methods still suffer from degenerate solutions and local minima when applied to long-period systems as briefly discussed in, for example, \citet{lebouquinResolvedAstrometricOrbits2017}.
In this context, we developed another approach, that of grid-based spectral disentangling, taking advantage of the very precise astrometric solution to drastically reduce the dimensionality of the parameter space that needs to be explored by spectral disentangling.\par

We base this work on the Fourier disentangling code \texttt{fd3} (formerly \texttt{FDBinary}; \citealt{ilijicObtainingNormalisedComponent2004}) by \citet{ilijicFd3SpectralDisentangling2017}.
Given the orbital elements $(e, P, T_0, \omega, K_1, K_2)$, \texttt{fd3} computes the separated spectra independently of any spectral templates.
Internally, this is achieved by solving a linear least-squares problem introduced by \citet{simonDisentanglingCompositeSpectra1994}, which was reformulated in Fourier space by \citet{hadravaOrbitalElementsMultiple1995}.
It is of the form
\begin{equation}\label{eq:linls}
	\tens{M}\vec{x} = \vec{c},
\end{equation}
where $\vec{x}= (\vec{x}^{\rm A} ~ \vec{x}^{\rm B})$ are row vectors representing the (unknown) separated component spectra $\vec{x}^{\rm A}$ and $\vec{x}^{\rm B}$ in Fourier space, $\vec{c}$ contains the Fourier components of the observed composite spectra, and $\tens{M}$ is the matrix that maps $\vec{x}$ to $\vec{c}$, which depends on the orbital solution of the system and times of observation.
The solver then uses the singular value decomposition of $\tens{M}$ as follows:
\begin{equation}\label{eq:svd}
	\tens{M} = \tens{U}\tens{\Sigma}\tens{V^{\rm T}},
\end{equation}
and calculates
\begin{equation}\label{eq:linlssol}
	\vec{x} = \tens{M}^{-1}\vec{c} = \tens{V}\tens{\Sigma}^{-1}\tens{U^{\rm T}}\vec{c}.
\end{equation} 
Since $\tens{M}$ generally has more rows than columns, the system is overdetermined and the least-squares solution $\vec{x_1}$ is returned, which minimizes the two-norm, and is given by
\begin{equation}\label{eq:linlsnorm}
	r = \left|\left|\frac{\tens{M}}{\vec{\sigma}}\vec{x_1} - \frac{\vec{c}}{\vec{\sigma}}\right|\right|_{2},
\end{equation}
which measures how well the orbital elements map the disentangled spectra $\vec{x_1}$ to the observed spectra $\vec{c}$, weighted with its measurement errors $\vec{\sigma}$.\par

Because Fourier transformations are applied, a disadvantage of this method is that the continuum level of the individual stars is undetermined with respect to the continuum of the composite spectra.
Not only must a flux ratio be supplied to set the individual contributions to the total flux, but numerical instabilities can create large undulations of the disentangled spectra.
Therefore, the output of the \texttt{fd3} algorithm must be renormalized again by fitting a smooth function through the regions, where
\begin{equation}
	\vec{x_1}^{\rm A} + \vec{x_1}^{\rm B} = 1.
\end{equation}

Our novel strategy is to fix the elements $(e, P, T_0, \omega)$ to those found from the astrometric orbit, and vary $K_1$ and $K_2$ in a grid of $K_1\in[15, 45]\,\kms{}$ and $K_2 \in [35, 65]\,\kms{}$ with $1\,\kms{}$ increments.
This grid is chosen with the expected total mass from the astrometric orbit and the mass ratio from \citet{rauwMultiwavelengthInvestigationNonthermal2002} in mind.
We decided to use a flux ratio equal to that observed in the H band, namely $f=f_{\rm H}=0.62$.
We recorded the reported reduced chi-squared distance $\chi^2 = r^2$ of Eq. \eqref{eq:linlsnorm} for each grid point, and these are represented in a contour map.
We wrapped the \texttt{C} code of \citet{ilijicFd3SpectralDisentangling2017} and the grid setup in a Python package available though GitHub\footnote{\url{https://github.com/matthiasfabry/gridfd3}}.
Earlier results of this disentangling strategy were presented on other challenging systems \citep{shenarHiddenCompanionLB12020, bodensteinerHR6819Triple2020}.\par

Instead of running the above procedure on the full wavelength domain of the spectra, which would be computationally expensive, we selected strong spectral lines and did the analysis on these lines separately.
We then summed the obtained chi-squared distances, and reduced them by dividing with the summed degrees of freedom, effectively clipping unwanted continuum regions.
The errors were estimated by a Monte Carlo (MC) sampling analysis, in which we drew 1500 samples of the full grid disentangling procedure, where the spectra are perturbed with Gaussian additive noise representative of their signal-to-noise ratio (S/N).
The orbital elements are perturbed with their 1$\sigma$ error given in Table \ref{tab:orbsol}.\par

The considered lines and wavelength ranges are shown in Table \ref{tab:fd3gridlines}. The criteria for selection are that they have high S/N, and that they are absorption lines.
Therefore, contrary to the analysis of \citet{rauwSagittariiUncoveringOtype2012}, weak metal lines like \ion{C}{iii} $\lambda$ 5696, \ion{Si}{iv} $\lambda$ 4116, \ion{O}{iii} $\lambda$ 5592 were neglected owing to their poor S/N with respect to the deep \element{H} and \ion{He}{ii} lines.\par

\begin{figure}
	\resizebox{\hsize}{!}{\includegraphics{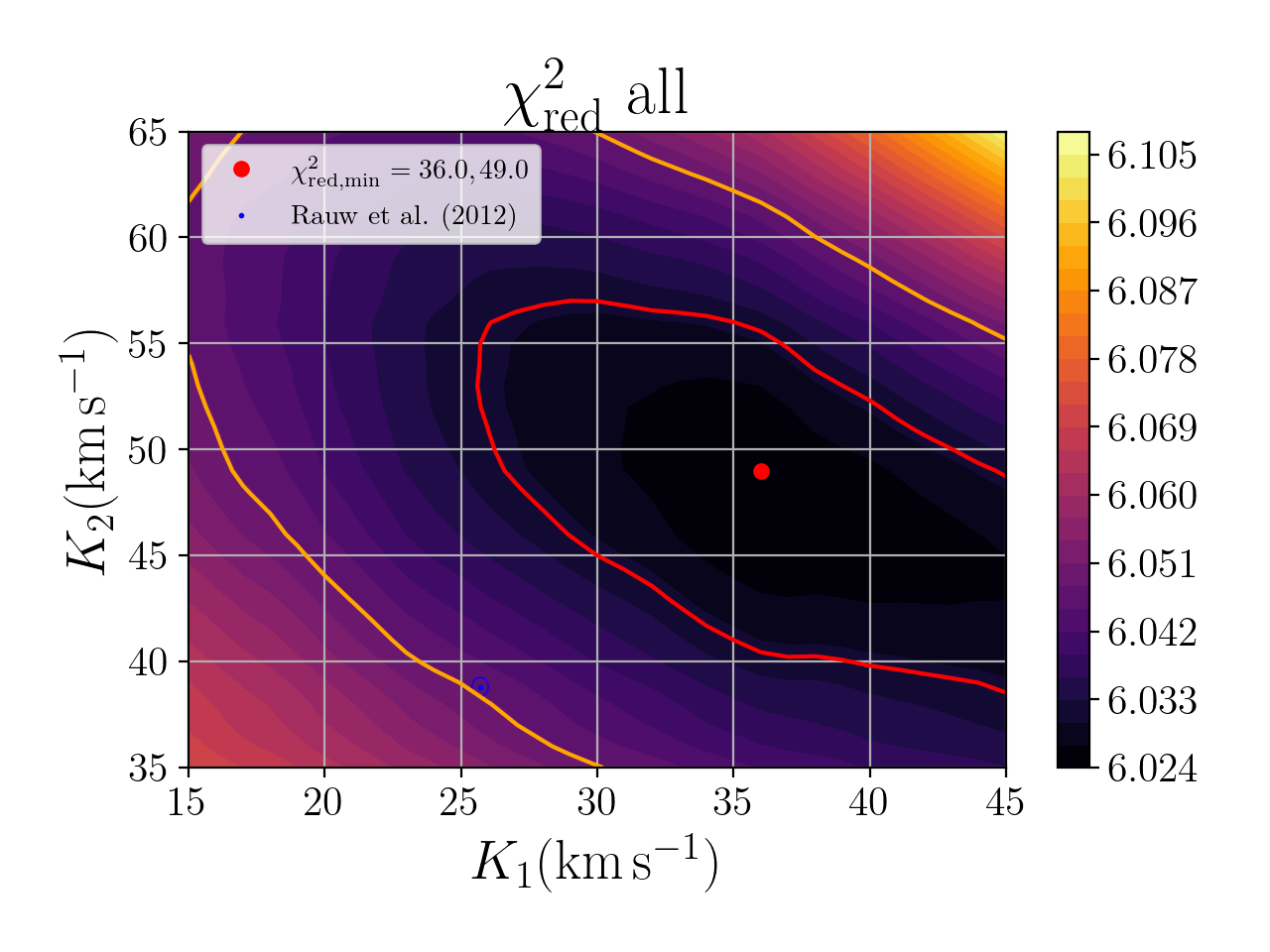}}
	\caption{Reduced chi-squared contour map of the grid disentangling combining all lines of Table \ref{tab:fd3gridlines}.
		The minimal $\chi^2_{\rm red}$, at $K_1 = 36\,\kms{}, K_2 = 49\,\kms{}$, is denoted with a red dot.
		The solid red and orange lines represent the 1$\sigma$ and 2$\sigma$ contours respectively, after scaling the $\chi^2$ map so that the minimum satisfies $\chi^2_{\rm red} = 1$.}
	\label{fig:chisq}
\end{figure}

Performing the \texttt{fd3} grid disentangling on the lines of Table \ref{tab:fd3gridlines} gives the reduced chi-squared contour plot in Fig. \ref{fig:chisq}.
We find that the $(K_1, K_2)$ pair that yields the minimal $\chi^2$ value is $K_1 = 36$~\kms{}, $K_2=49$~\kms{} and lies in a rather shallow minimum.
We used the difference of the (marginalized) 0.84 and 0.16 percentiles of the MC sampling as estimators of the 1$\sigma$ confidence interval, yielding  $K_1 = 36^{+4}_{-1}$~\kms{}, $K_2=49\pm3$~\kms{}, resulting in dynamical masses of $53^{+7}_{-6}\Msun{}$ and $39^{+6}_{-3}\Msun{}$ for the primary and secondary respectively.
This differs to the values of \citet{rauwSagittariiUncoveringOtype2012}, who measured $K_1 = 25.7\pm0.6\,\kms{}$ and $K_2 = 38.8\pm0.9\,\kms{}$ and needed to invoke a much lower inclination to reconcile these with the inferred masses from their spectral types.
While the mass ratio $K_1/K_2$ is similar to our results, the semi-amplitude values from \citet{rauwSagittariiUncoveringOtype2012, rauwTestingTheoryColliding2016} differ by 30\%, that is, more than the statistical uncertainties of the fit.
In this sense, in this work we present a different orbital solution.
\citet{rauwSagittariiUncoveringOtype2012, rauwTestingTheoryColliding2016} use the mean of the RVs from the \ion{Si}{iv} $\lambda$ 4116 and \ion{N}{v} $\ll$ 4604, 4620 lines for the primary and similarly the mean from the \ion{He}{i} $\ll$ 4471, 5876, \ion{O}{iii} $\lambda$ 5592, \ion{C}{iii} $\lambda$ 5696, and \ion{C}{iv} $\ll$ 5801, 5812 lines for the secondary.
The authors argue that while most of these lines are in fact SB2 lines (as visible near maximum RV separation), the line blend is dominated by only one component such that they treated them as SB1 lines tracing only the motion of that component.
The spectral disentangling results however show that both components contribute to these lines, sometimes significantly.
For example, Fig~\ref{fig:reconoiii5592} shows the \ion{O}{iii} $\lambda$ 5592 line, to which the primary contributes about 25\% as compared to the total depth of the line.
As discussed in \citet{sanaNonthermalRadioEmitter2011} and quantitatively investigated in \citet[][in press]{bodensteinerYoungMassiveSMC2021}, the effect may be more subtle than this example because an even smaller contamination impacts the measured RVs.
While for many systems this is of minor importance, in the present case, even a bias as small as $5\,\kms{}$ on the RVs is large (${\sim}10\%$) compared to the measured semi-amplitudes.\par

\begin{figure}
	\resizebox{\hsize}{!}{\includegraphics{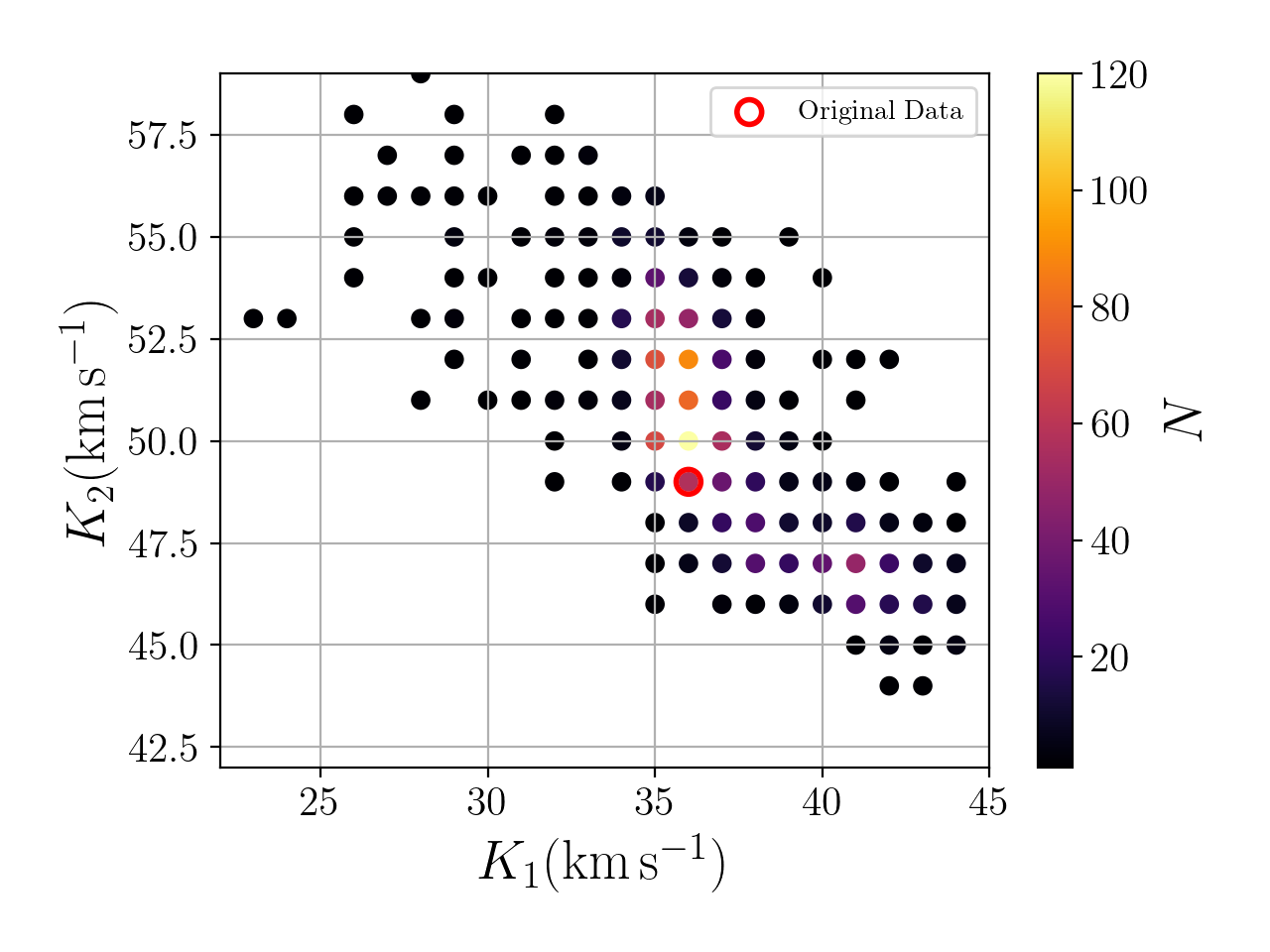}}
	\caption{Two-dimensional histogram of the results of the MC sampling in $(K_1, K_2)$ space.
		The color indicates the number of samples $N$ at the corresponding $(K_1, K_2)$ pairs.
		The result from the original data is encircled in red.}
	\label{fig:hist}
\end{figure}

The results of the \texttt{fd3} grid search are summarized in Table~\ref{tab:k1k2mc}.
The error analysis shows $K_1$ has a slightly asymmetric 1$\sigma$ uncertainty, as indicated in Table~\ref{tab:k1k2mc} and shown in the 2D $(K_1, K_2)$ histogram in Fig.~\ref{fig:hist}.
Yet, for both $K_1$ and $K_2$, the median of the retrieved MC values lies within 1$\sigma$ of the input value, giving us confidence in the accuracy of the method. For comparison, the marginalized histograms are shown in Fig.~\ref{fig:k1k2marg}.\par

\begin{table}
	\caption{Results of the \texttt{fd3} grid search for the RV semi-amplitudes. Column two gives the grid point that has the minimal reduced chi-squared value when disentangling the original data. Columns 3, 4 and 5 give the MC median and 1$\sigma$ confidence intervals.}
	\label{tab:k1k2mc}
	\centering
	\begin{tabular}{c | c | c c c}
		\hline\hline
		Semi-amplitudes & Data & MC median & $\sigma_-$ & $\sigma_+$ \\
		\hline
		$K_1(\kms{})$ & 36 & 36 & 35 & 40\\
		$K_2(\kms{})$ & 49 & 50 & 47 & 53\\
		\hline
	\end{tabular}
\end{table}

\begin{figure*}
	\resizebox{\hsize}{!}{\includegraphics{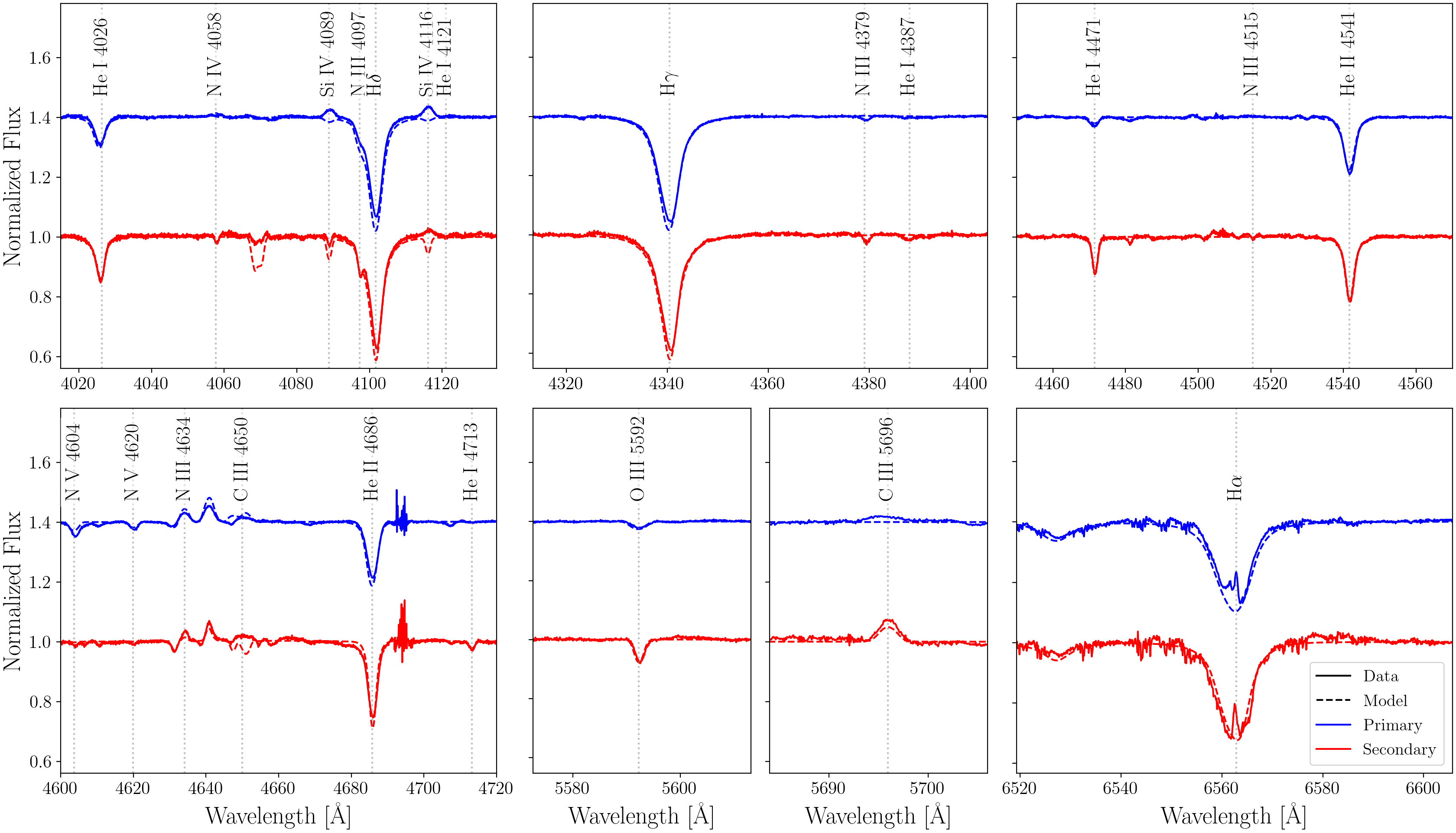}}
	\caption{Disentangled, renormalized spectra of both components using $K_1 = 36\,\kms{}$ and $K_2 = 49\,\kms{}$. The spectrum of the primary  is shifted up by 0.4 units with respect to the secondary for clarity. In \halpha{}, the core is contaminated by nebular emission present in the FEROS and HERMES spectra. The model spectra are those obtained with the best-fit atmospheric parameters from the \textsc{fastwind} analysis (Sect. \ref{sec:atm}).}
	\label{fig:disenspectra}
\end{figure*}

We show the broad range disentangled spectra in Fig. \ref{fig:disenspectra}.
From visual inspection of the line depth ratio of \ion{He}{i} $\lambda$ 4471 to \ion{He}{ii} $\lambda$ 4541, we immediately notice that the primary is the hottest star of the two.
This fact is seen in the \ion{N}{v} $\ll{}$ 4604, 4620 lines as well.
Comparing equivalent widths of lines in the disentangled spectra with classification tables of \citet{contiSpectroscopicStudiesOType1971} and Sana et al. \citetext{in prep.}, we update the classification of the primary by \citet{rauwSagittariiUncoveringOtype2012} from O3.5V((f$^+$)) to O3V((f$^+$)), and narrow down the subtype of the secondary from O5-5.5V((f)) to O5V((f)). This is in excellent agreement with the recent work of \citet{quinteroMassiveBinarySystem2020}, who used an updated version of the shift-and-add disentangling technique in wavelength space and also obtained spectral types of O3V((f$^+$)) and O5V((f)).\par

To confirm the quality of the disentangling procedure, we recombine the disentangled spectra at various phases and compute residuals.
To avoid mixing different instrumental wavelength calibrations, only the HERMES spectra are used. Reconstructed lines are shown in Figs. \ref{fig:reconhdelta} and \ref{fig:reconoiii5592}.
The average residuals of the reconstructed lines to the original composite spectra are on the order 1\%.
While very good, this is larger than expected from pure S/N considerations and probably results from small uncertainties in the normalization of individual spectra that impacts the disentangling results of \texttt{fd3}.\par

\subsection{Distance}\label{sec:dis}
Because the semi-amplitudes $K_1$ and $K_2$ set the absolute scale of the orbit, we can infer the distance to \objectname{} by measuring against the apparent orbit found through the interferometry (Fig.~\ref{fig:asfig}), via the total mass of the system and Kepler's third law, as follows:
\begin{equation}
	d = \frac{1}{a_{\rm app}}\sqrt[\leftroot{-1}\uproot{2}\scriptstyle 3]{\frac{G(M_1+M_2)P^2}{4\pi^2}},
\end{equation}
where $a_{\rm app}$ is the semimajor axis of the apparent orbit in angular units.
The values $(K_1, K_2) = (36, 49)\,\kms{}$ correspond to a total dynamical mass of $92\,\Msun{}$, which results in a distance of
\begin{equation}\label{eq:dist}
	d = 1310 \pm 60 \,\si{pc},
\end{equation}
where we propagated the MC errors.

This value lies within the Gaia eDR3 geometric distance reported by \citet{bailer-jonesEstimatingDistancesParallaxes2020} of $1218^{+100}_{-108}\,\si{pc}$.
With these results, it seems likely that \objectname{} is a member of NGC6530, confirming the measurements of \citet{prisinzanoStarFormationRegion2005} and especially \citet{kharchenkoAstrophysicalParametersGalactic2005}, who quoted a distance of $d=1322\,\si{pc}$.
Conversely, we can firmly exclude distances to \objectname{} of over $1500\,\si{pc}$ because that would require a total mass of over $130\Msun{}$, which our grid disentangling results do not support.
Similarly, the distance of $1780\pm 80\,\si{pc}$ of \citet{sungUBVRIHaPhotometry2000} adopted by \citet{rauwSagittariiUncoveringOtype2012} is incompatible with the interferometric and Gaia measurements of the apparent size of the orbit and the derived component masses.\par
\section{Atmosphere modeling}\label{sec:atm}
\subsection{Setup}
Using the disentangled spectra, we adjust theoretical line profiles computed with the \textsc{fastwind} NLTE atmosphere code suitable for the expanding atmosphere of O-type stars  \citep{santolaya-reyAtmosphericNLTEmodelsSpectroscopic1997, pulsAtmosphericNLTEmodelsSpectroscopic2005, carneiroAtmosphericNLTEModels2016, pulsFASTWINDReloadedComplete2017, sundqvistAtmosphericNLTEModels2018}.
To reduce the dimensionality of the parameter space, the rotational and macroturbulent velocities are estimated using the \texttt{iacob-broad} tool \citep{simon-diazIACOBProjectRotational2014}.
Analyzing (with the goodness-of-fit method) the \ion{O}{iii} $\lambda$ 5592 and \ion{N}{V} $\lambda$ 4603 lines for the primary and \ion{O}{iii} $\lambda$ 5592, \ion{He}{i} $\lambda$ 4713 and \ion{He}{i} $\lambda$ 5876 for the secondary, we find a projected rotational velocity $v\sin i=102^{+8}_{-12}\,\kms{}$ and a macroturbulent velocity $v_{\rm mac}=77^{+23}_{-20}\,\kms{}$ for the primary star, while we take $v\sin i=67^{+6}_{-13}\,\kms{}$ and $v_{\rm mac}=48^{+21}_{-14}\,\kms{}$ for the secondary.
Similar values are obtained when using the Fourier transform method.\par

The stellar atmosphere models are then iterated using a genetic algorithm  \citep{charbonneauGeneticAlgorithmsAstronomy1995,mokiemSpectralAnalysisEarlytype2005} within a predefined parameter space to optimize a $\chi^2$ fitness metric until a convergence to the best fit with the spectrum is reached.
The version of the genetic algorithm used is detailed in \citet{abdul-masihCluesOriginEvolution2019}.
We set the $\beta$ exponent of the wind acceleration law  to 0.85, as appropriate for main-sequence stars \citep{muijresPredictionsMasslossRates2012}.
The microturbulence velocity is fixed to  $v_{\rm mic}=10\,\kms{}$ in the computation by \textsc{fastwind}; in the formal integral it is selected on the criteria $v_{\rm mic} = \max(10\,\kms{}, 0.1v_\mathrm{wind})$.
We include optically thin wind clumping, with a near constant clumping factor $f_{\rm cl}$ throughout the wind.
Lastly, we opted to clip the core of the \halpha{} line to avoid fitting the nebular emission remnant (visible in the bottom-rightmost panel of Fig. \ref{fig:disenspectra}).
The full list of fitted spectral lines is shown in Table \ref{tab:atmlines}. \par

To provide an absolute magnitude anchor point for the atmospheric model, we adopt the photometric data from the Two Micron All-Sky Survey (2MASS, \citealt{skrutskieTwoMicronAll2006}), giving an apparent magnitude in the near-infrared K${}_{\rm S}$ band of $m_{\rm K_S} = 5.731 \pm 0.024$ for the total system.
Using the interstellar absorption coefficient $A_V = 1.338 \pm 0.021$ from \citet{maizapellanizOpticalNIRDustExtinction2018}, the color correction $A_{\rm K_S}/A_V = 0.116$ from \citet{fitzpatrickCorrectingEffectsInterstellar1999} and a distance of $1.31\pm0.06\,\si{kpc}$ calculated in Sect.~\ref{sec:dis} results in an absolute magnitude in the K${}_{\rm S}$ band of $M_{\rm K_S} = -5.01\pm0.10$.
We correct for the measured PIONIER flux ratio $f=0.62$ between the components, where we assume that it remains unchanged between the H and K${}_{\rm S}$ band (as expected for such hot objects).
This yields absolute component magnitudes of $M_{\rm K_S} = -4.49 \pm 0.10$ and $-3.96 \pm 0.10$ for the primary and secondary, respectively.
We note that these values correspond to fainter stars than their derived spectral types suggest.
Synthetic photometry of \citet{martinsUBVJHKSyntheticPhotometry2006} give $M_{\rm K} = -4.98$ and $M_{\rm K} = -4.39$ for the O3V primary and O5V secondary, respectively, which suggests the stars are slightly more compact.\par

\subsection{Results and discussion}\label{sec:atmres}
We list the spectroscopic parameters of the resulting best-fit atmospheric models and the resulting inferred parameters in the leftmost column of Table~\ref{tab:spec_phot_evol}.
The corresponding theoretical spectra are plotted in Fig.~\ref{fig:disenspectra} with dashed lines.
We note the obvious nebular contamination of \halpha{}, as well as the general trend that the disentangled spectra are slightly shallower than the model spectra in the deep and broad lines (like \hdelta{}, \hgamma{} and \ion{He}{ii} $\lambda$ 4686), suggesting issues in the normalization of these broad lines. \par

The best-fit parameters depend on which line features were considered in the fit and with what weights.
For example, giving more weight to \ion{He}{i} lines would result in lower inferred effective temperatures and vice versa.
Correspondingly, the inferred surface gravities would be lower for lower $\teff{}$ and vice versa.
Therefore, conservative errors of $1\si{\kilo\kelvin}$ and $0.2\,\si{dex}$ on $\teff{}$ and $\log g$, respectively, are adopted.
Furthermore, the determination of the quality of the CNO abundance measurements is challenging.
The carbon abundance for the secondary, for example, is fitted to $\C{} = 9.12$; this is an unusually high measurement that has (to our knowledge) never before been observed.
We note that this measurement is driven by the \ion{C}{iii} $\lambda$ 5696 line, which is in emission.
At $\teff{} = 42 \si{\kilo\kelvin}$, \textsc{fastwind} can only reconcile this line in emission by boosting the carbon abundance.
Keeping the issues presented by this line in mind, however \citep[see][]{martinsFormationIii464750512012},  we adopt the minimum of a formal $0.1\,\si{dex}$ error and the statistical error of the grid of models.
Since this formal error is somewhat arbitrary, even these uncertainties should be interpreted with great care.
Therefore we can only argue for qualitative enrichment of nitrogen in the primary and enrichment of carbon in the secondary.
For added justification, we show in Figs.~\ref{fig:abunprim} and \ref{fig:abunsec} the comparison of the model spectra, their error ranges along with spectra using the \citet{brottRotatingMassiveMainsequence2011} CNO baseline abundances in various diagnostic lines of the CNO elements.\par

The best-fit $\log g$ values then provide the spectroscopic masses of the stars, which are found to be $32\pm16\Msun{}$ and $19\pm10\Msun{}$ for the primary and secondary, respectively.
These masses are significantly lower than their dynamical counterparts, albeit with large error bars, and are not representative of dwarf stars of that luminosity.
The main reason for this discrepancy is the low inferred $\log g$, which should be raised by about $0.25\,\si{dex}$ for both stars, that is, slightly beyond the adopted uncertainty, to match the dynamical masses.
The mass discrepancy problem \citep{herreroIntrinsicParametersGalactic1992} is still an open issue in massive-star spectroscopy, and while more recent studies \citep[\textit{e.g.,}][]{mahyTarantulaMassiveBinary2020} show that for stars above ${\sim}35\Msun{}$, the discrepancy largely disappears, in this analysis, it is still present.
The repeated normalization of the spectra before and after disentangling could be the root cause of this fact, as the reconstruction plots in Fig.~\ref{fig:reconhdelta} and \ref{fig:reconoiii5592} and the comparison to the model spectra (Fig.~\ref{fig:disenspectra}) hint towards.

\section{Evolutionary modeling}\label{sec:evol}
We compare our previous results with the Milky Way evolutionary tracks of \citet{brottRotatingMassiveMainsequence2011}, using the Bayesian search tool \textsc{bonnsai}\footnote{The BONNSAI web-service is available at www.astro.uni-bonn.de/stars/bonnsai.} \citep{schneiderBonnsaiBayesianTool2014}.
The \textsc{bonnsai} tool allows us to search the rotating single star evolution tracks of \citet{brottRotatingMassiveMainsequence2011} for the highest likelihood stellar model that corresponds to measured quantities.
We input the observed $\log L, \teff{}, X_{\rm He}$, and $v\sin i$ from the left column of Table \ref{tab:spec_phot_evol} and request the highest likelihood models of both stars in the grid.
To avoid biasing the Bayesian search, we refrain from using the $\log g$ due to the uncertainties posed by the mass discrepancy.
In a first search, we do not input the CNO abundances obtained from \textsc{fastwind}.
The parameters from the highest likelihood models replicated from our spectroscopic and photometric observables of this search are given in the middle column of Table \ref{tab:spec_phot_evol}.\par

\begin{table*}
	\caption{Parameters of the best-fit, genetically evolved \textsc{fastwind} atmospheric model (described in Sect.~\ref{sec:atm}), along with replicated observables from the \citet{brottRotatingMassiveMainsequence2011} models using \textsc{bonnsai} (Sect.~\ref{sec:evol}).
		The errors correspond to the 1$\sigma$ confidence level. Empty entries are indeterminable for that parameter.}
	\label{tab:spec_phot_evol}
	\centering
	\begin{tabular}{c | c c | c c | c}
		\hline\hline
		& \multicolumn{2}{c|}{\textsc{fastwind}} & \multicolumn{3}{c}{\textsc{bonnsai}}\\
		\hline
		Parameter(Unit) & Primary & Secondary & Primary & Secondary & Primary, Scaled CNO \\
		\hline
		$\teff{}\,[\si{k\K}]$ & $46.0\pm1.0$ & $42.0\pm1.0$& $45.9^{+0.6}_{-0.9}$ & $41.9\pm0.9$  & $46.0^{+0.6}_{-1.0}$\\
		$\log(g/{\rm [cgs]})$  & $3.87\pm0.20$ & $3.87\pm0.20$ & $4.11\pm0.05$ & $4.12^{+0.06}_{-0.07}$ & $4.10^{+0.02}_{-0.06}$\\
		$\log\frac{\Dot{M}}{\Msun{}/\si{yr}}$ & $-6.6\pm0.2$ & $-6.6\pm0.2$ &\ldots&\ldots&\ldots\\
		$f_{\rm cl}$ & $29\pm 5$ & $22\pm 3$ &\ldots&\ldots&\ldots \\
		$v\sin i\, [\kms{}]\tablefootmark{a}$ & $102^{+8}_{-12}$ & $67^{+6}_{-13}$ &\ldots &\ldots &\ldots\\
		$v_{\rm rot}\,[\kms{}]$ & \ldots& \ldots& $110^{+59}_{-26}$ & $70^{+8}_{-15}$  & $330^{+26}_{-30}$\\
		\hline
		$Y_{\rm He}$ & $0.25\pm0.04$ & $0.24\pm0.03$  & $0.26\tablefootmark{b}$ & $0.26\tablefootmark{b}$  & $0.28^{+0.08}_{-0.02}$\\
		{[C/H]} + 12 & $8.17^{+0.60}_{-0.55}$ & $9.12\pm0.10$\tablefootmark{*}  & $8.14^{+0.01}_{-0.03}$ & $8.13\tablefootmark{b}$  & $7.12^{+0.55}_{-0.05}$ \\
		{[N/H]} + 12 & $8.45^{+0.10}_{-0.29}$ & $7.42\pm0.10$  & $7.63^{+0.09}_{-0.01}$ & $7.64\tablefootmark{b}$& $8.72^{+0.10}_{-0.27}$\\
		{[O/H]} + 12 & $8.63^{+0.10}_{-0.70}$ & $8.64^{+0.10}_{-0.13}$ & $8.55^{+0.01}_{-0.02}$ & $8.55\tablefootmark{b}$ & $8.55^{+0.01}_{-0.61}$\\
		\hline
		$\log(L/\Lsun{})$ & $5.68 \pm 0.08$ & $5.35 \pm 0.08$ & $5.64^{+0.07}_{-0.06}$ & $5.33^{+0.08}_{-0.06}$ & $5.67^{+0.06}_{-0.07}$\\
		$R\,[\Rsun{}]$ & $10.8\pm1.0$& $8.9\pm1.2$ & $10.45^{+0.88}_{-0.59}$ & $8.73^{+0.75}_{-0.67}$ & $10.73^{+0.79}_{-0.61}$\\
		$M_{\rm spec}\,[\Msun{}]$ & $32.1\pm16.0$& $18.9\pm10.1$ &\ldots&\ldots&\ldots\\
		$M_{\rm evol}\,[\Msun{}]$ &\ldots&\ldots& $53.4^{+3.2}_{-3.3}$& $37.0^{+2.0}_{-2.3}$& $53.8\pm4.7$\\
		Age $[\si{Myr}]$ &\ldots&\ldots& $0.52^{+0.32}_{-0.33}$ & $1.00^{+0.48}_{-0.58}$ & $1.00^{+0.80}_{-0.41}$\\
		\hline
	\end{tabular}
	\tablefoot{
		\tablefoottext{a}{Determined using \textsc{iacob-broad}, not \textsc{fastwind}.}
		\tablefoottext{b}{Very small error, see Sect.~\ref{sec:evol}.}
		\tablefoottext{*}{Highly uncertain measurement, this formal error is likely not representative, see Sect.~\ref{sec:atmres}.}
	}
\end{table*}

The comparison with evolutionary tracks point toward relatively compact and coeval stars with an age of about $1~\si{\mega yr}$.
We find that the evolutionary masses are within error of the dynamical masses, which provides a further indication that the spectroscopic mass likely suffers from systematic errors.
Additionally, the evolutionary models favor lower CNO surface abundances than are spectroscopically inferred, especially nitrogen in the primary and carbon in the secondary.
The rather modest rotational velocities and young ages do not allow for rotational mixing to modify the surface composition; the CNO abundances returned by \textsc{bonnsai} correspond to the baseline value of the \citet{brottRotatingMassiveMainsequence2011} models with very small uncertainties.\par

From the atmosphere models in Table~\ref{tab:spec_phot_evol}, it is clear that the CNO composition between the primary and the secondary is different.
This is not reflected in the evolutionary tracks because both models prefer the baseline values of the \citet{brottRotatingMassiveMainsequence2011} tracks, namely $\C{} = 8.13, \N{} = 7.64$ and $\O{}=8.55$ (middle column of Table~\ref{tab:spec_phot_evol}).
Furthermore, if we assume the abundances of the secondary are baseline for \objectname{}, this source has a different CNO baseline than the \citet{brottRotatingMassiveMainsequence2011} tracks.
It is hard to justify observationally that the observed abundances of the secondary are baseline for \objectname{} or the cluster NGC6530.
But we expect the least massive star with the lowest rotational velocity to be least contaminated by surface enrichment from a theoretical standpoint.
We thus test if \textsc{bonnsai} finds different models for the primary if we scale down the observed CNO abundances to the \citet{brottRotatingMassiveMainsequence2011} baseline.
For the abundances of the primary to maintain the same fractional difference versus the secondary, this amounts to calculating $[X'/{\rm H}] = [X/{\rm H}]_{\rm base,Brott} - [X/{\rm H}]_{\rm base,obs} + [X/{\rm H}]_{\rm prim,obs}$.
Keeping the doubtful C abundance measurement of the secondary in mind (Sect.~\ref{sec:atmres}), we refrain from scaling C and compute $\N{'} = 8.67^{+0.14}_{-0.31}$ and $\O{'} = 8.54^{+0.14}_{-0.71}$.
Using then again $\log L, \teff{}, X_{\rm He}$, and $v\sin i$ from the \textsc{fastwind} models, along with these new N and O abundances as input for \textsc{bonnsai}, we obtain other highest likelihood evolutionary parameters; these are listed in the rightmost column of Table~\ref{tab:spec_phot_evol}.
These results point to a different scenario.
Here the rotational velocity is significantly higher, allowing significant rotational mixing to occur.
The primary age has increased to match that of the secondary as well, while the evolutionary mass, $\log g$ and $\teff{}$ are only slightly changed when comparing to the nonscaled results (middle column of Table~\ref{tab:spec_phot_evol}).
The major implication of this is that either the rotational axis and the normal to orbital plane are heavily inclined, up to an estimated 68 degrees to explain the observed projected rotational velocity, or the effect of rotational mixing is underestimated in the stellar evolution models.
We cannot exclude either that a mixing mechanism weakly dependent on rotation is operating on stars in this mass regime.
Distinguishing between these scenarios however requires greater confidence in the quality of the CNO abundance measurements. \par

We summarize our results of Sects.~\ref{sec:atm} and \ref{sec:evol} in an HRD and a Kiel diagram that is overplotted on several of the evolutionary tracks and isochrones of \citet{brottRotatingMassiveMainsequence2011}.
We note that while the location of the models on the HR diagram matches well, there is a mismatch of the spectroscopic mass inferred from the \textsc{fastwind} models and the evolutionary masses.
In the Kiel diagram, there is a poorer match as expected from the mass discrepancy discussed in Sect.~\ref{sec:atmres}.\par

\begin{figure}
	\centering
	\includegraphics[width=\columnwidth]{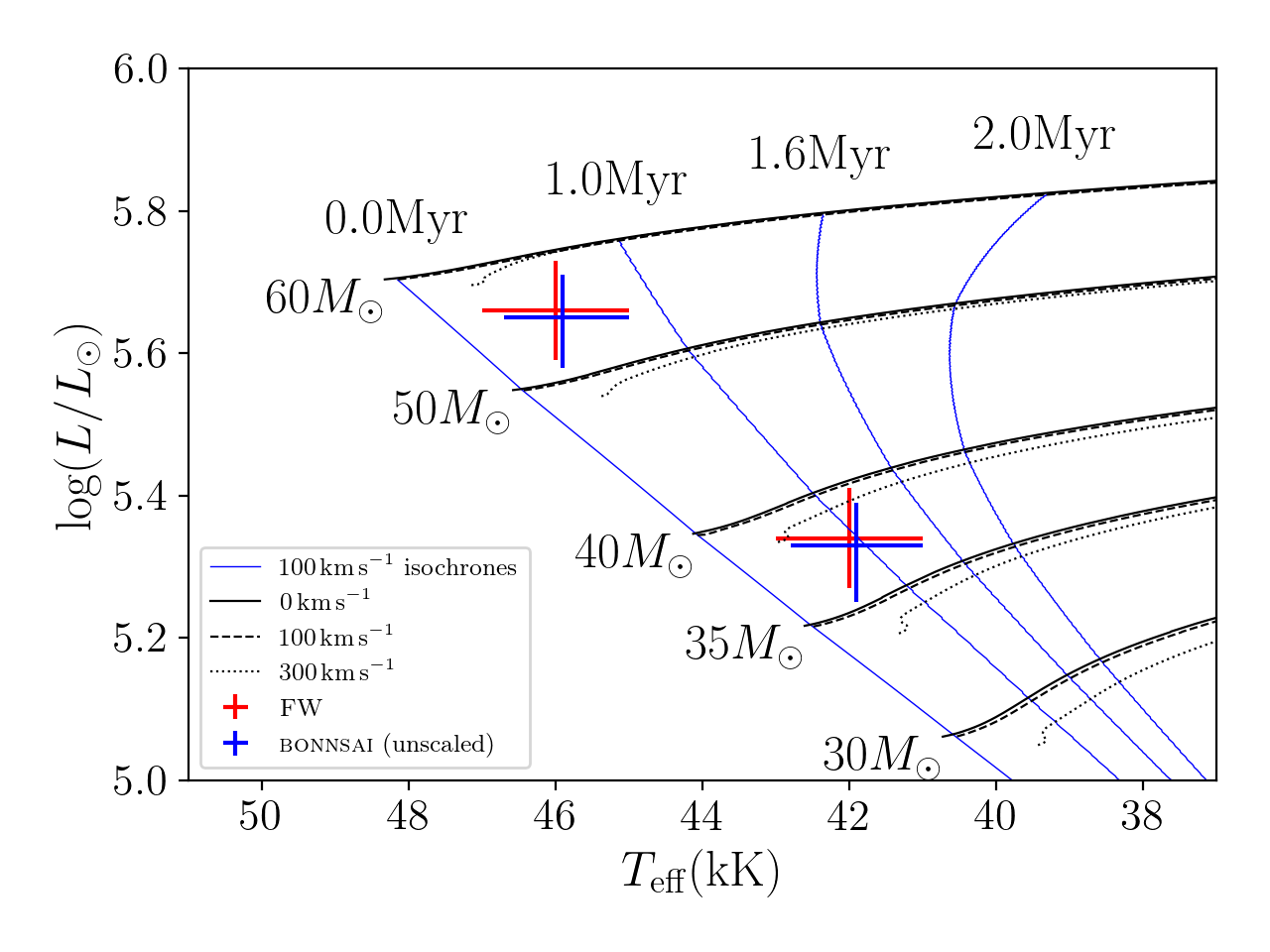}
	\includegraphics[width=\columnwidth]{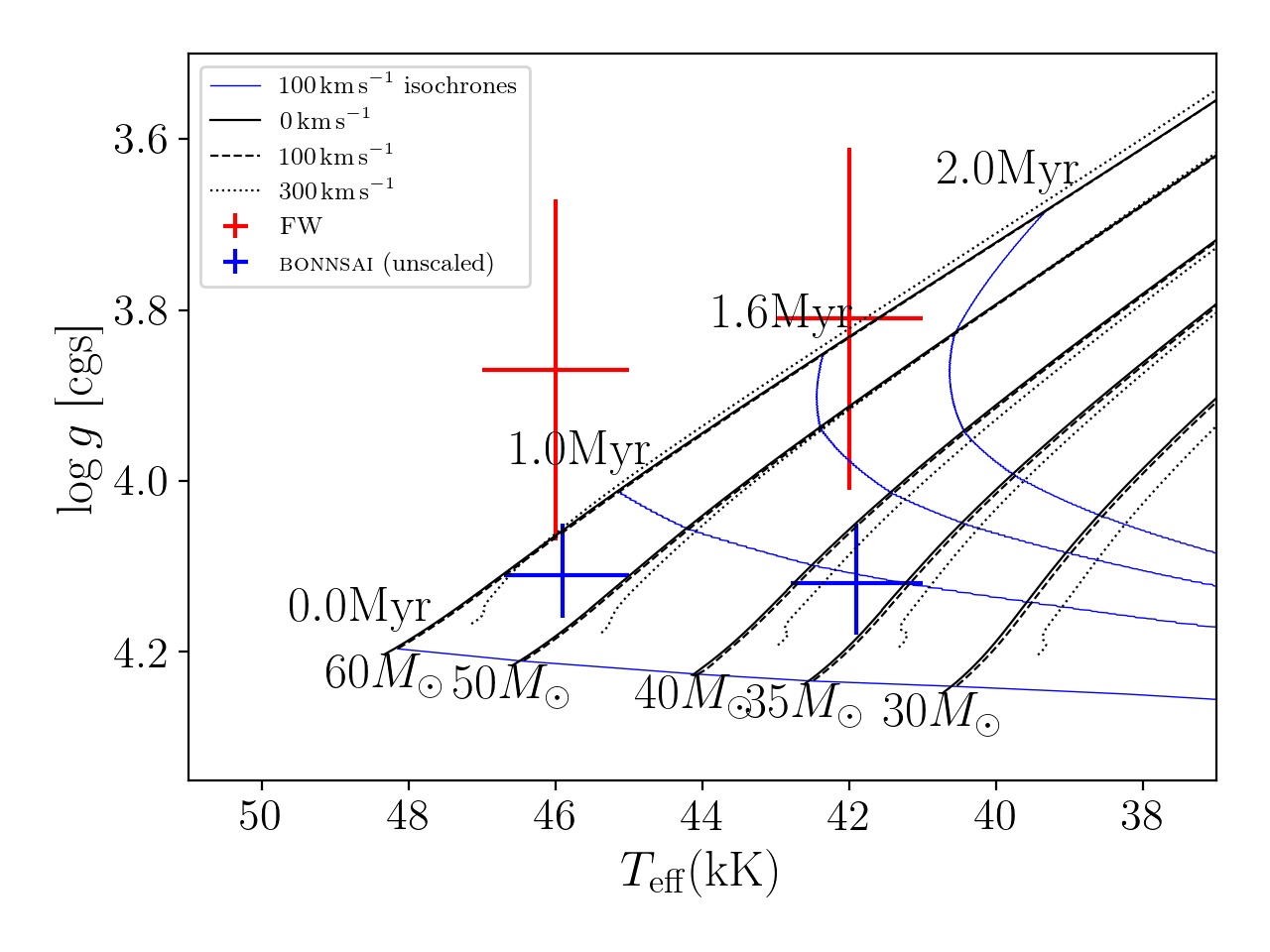}
	\caption{{\it Top:} Hertzsprung-Russell diagram, showing the location of the best-fit \textsc{fastwind} models (FW) and highest likelihood evolutionary models from \citet{brottRotatingMassiveMainsequence2011} (\textsc{bonnsai}).
		Overplotted are evolutionary tracks for different masses and initial rotation velocities (grey lines), along with isochrones for the $100 \kms{}$ initial rotation velocity models (blue lines).
		{\it Bottom:} Kiel diagram with an equivalent legend as the HRD above.
		Using the scaled CNO abundances does not appreciably move the model of the primary in either diagram.}
	\label{fig:my_label}
\end{figure}
\section{Conclusions}\label{sec:conc}
We have obtained disentangled spectra of \objectname{} using a combination of high angular resolution astrometry and spectral grid disentangling with the \texttt{fd3} code.
The astrometric measurements solidify the long period of 8.9 yr and have a near edge-on inclination of 86.5 degrees.
Our results confirm the presence of an O3V+O5V massive binary, which has inferred dynamical masses of about 53 and 39 $\Msun{}$, making \objectname{} one of the most massive galactic O+O binaries ever resolved.
Furthermore, to our knowledge, this is only the second instance of a dynamical mass estimate of a galactic O3V star (the other from \citealt{mahyTripleSystemHD2018}).\par

By re-deriving the semi-amplitudes of the RV curves, we clear up the contradictory results between the previous RV measurements of \citet{rauwSagittariiUncoveringOtype2012} and the high inclination of the interferometric orbit by \citet{lebouquinResolvedAstrometricOrbits2017}.
Furthermore, the results show \objectname{} is a member of the young open cluster NGC~6530.
The combined dynamical, atmospheric, and evolutionary modeling shows \objectname{} contains massive stars of roughly $53 \Msun{}$ and $37\Msun{}$ for the primary and secondary, respectively.
\objectname{} is a unique system in the far top left corner of the HRD, and therefore provides an equally unique opportunity to use its stellar and systemic parameters to compare with massive star evolutionary models as well as binary formation scenarios.\par

\begin{acknowledgements}
	This project has received funding from the European Research Council (ERC) under the European Union's Horizon 2020 research and innovation programme (grant agreement number DLV-772225-MULTIPLES), the  KU Leuven Research Council (grant C16/17/007: MAESTRO), the FWO through a FWO junior postdoctoral fellowship (No. 12ZY520N) as well as the European Space Agency (ESA) through the Belgian Federal Science Policy Office (BELSPO). Based on observations obtained with the HERMES spectrograph, which is supported by the Research Foundation - Flanders (FWO), Belgium, the Research Council of KU Leuven, Belgium, the Fonds National de la Recherche Scientifique (F.R.S.-FNRS), Belgium, the Royal Observatory of Belgium, the Observatoire de Genève, Switzerland and the Thüringer Landessternwarte Tautenburg, Germany.
\end{acknowledgements}

\bibliographystyle{bibtex/aa}
\bibliography{bibtex/Massive_Stars.bib}

\appendix
\onecolumn
\section{Additional tables and figures} \label{app:figs}
\begin{table}[h]
	\caption{Previously unanalyzed spectra of \objectname{}. The columns signify the modified Julian date (MJD) of observation, spectrograph, exposure time and estimated S/N at $4500\si{\angstrom}$. }
	\label{tab:newspectra}
	\centering
	\begin{tabular}{c c c c}
		\hline\hline
		MJD & Instrument & exposure time (\si{\second}) & S/N\\
		\hline
		57171.1874 & HERMES & 571 & 90\\
		57172.1426 & HERMES & 325 & 96\\
		57173.1646 & HERMES & 478 & 75\\
		57201.0247 & HERMES & 360 & 138\\
		57236.9486 & HERMES & 800 & 161\\
		57634.0119 & HARPS  & 180 & 143\\
		\hline
	\end{tabular}
\end{table}

\begin{table}[h]
	\caption{Journal of the interferometric measurements of \objectname{}.
		Given are the VLTI instrument (Instr.), the MJD of observation, the relative angular separation (Sep.), the position angle (PA) east of north, and the major ($\sigma_{\text{maj}}$) and minor ($\sigma_{\text{min}}$) axes of the 1$\sigma$ error ellipses along with its PA east of north (PA${}_{\sigma}$).}
	\label{tab:newinter}
	\centering
	\begin{tabular}{c c c c c c c}
		\hline\hline
		Instr. & MJD & Sep. (\si{mas}) & PA (\si{deg}) & $\sigma_{\text{maj}}$(\si{mas}) & $\sigma_{\text{min}}$(\si{mas}) & PA${}_{\sigma}(\si{deg})$ \\
		\hline
		AMBER & 54995.306 & 20.74 & 69.87 &  1.12 &  0.75 &  154 \\
		AMBER & 55644.304 & 11.85 & 71.58 &  1.20 &  0.50 &  116 \\
		AMBER & 55648.389 & 11.76 & 72.19 &  0.83 &  0.63 &  113 \\
		PIONIER & 56154.134 &  0.74 & 149.10 &  1.85 &  0.82 &   16 \\
		PIONIER & 56189.027 &  1.09 & 207.39 &  1.03 &  0.48 &  153 \\
		PIONIER & 56191.087 &  1.05 & 201.56 &  0.57 &  0.21 &  138 \\
		PIONIER & 56376.318 &  4.87 & 242.59 &  0.50 &  0.23 &   44 \\
		PIONIER & 56383.324 &  4.94 & 241.83 &  1.82 &  0.31 &  112 \\
		PIONIER & 56438.244 &  5.57 & 244.91 &  0.13 &  0.06 &  151 \\
		PIONIER & 56549.143 &  4.40 & 247.37 &  2.12 &  0.36 &  141 \\
		PIONIER & 56714.359 &  3.23 & 59.89 &  0.21 &  0.18 &  150 \\
		PIONIER & 56751.389 &  4.92 & 62.51 &  0.24 &  0.10 &  135 \\
		PIONIER & 56783.350 &  6.23 & 63.05 &  0.22 &  0.09 &  121 \\
		PIONIER & 56787.340 &  6.43 & 63.33 &  0.12 &  0.09 &  127 \\
		PIONIER & 56818.282 &  7.56 & 64.46 &  0.50 &  0.23 &  137 \\
		PIONIER & 56903.045 & 10.63 & 65.49 &  0.25 &  0.22 &   6 \\
		PIONIER & 57558.303 & 21.25 & 67.89 &  0.43 &  0.24 &  176 \\
		GRAVITY & 57558.305 & 21.194 & 68.028 & 0.101 & 0.102 & 90 \\
		GRAVITY & 57560.312 & 21.45 & 67.90 & 0.24 & 0.22 & 90 \\
		GRAVITY & 57647.081 & 21.664 & 68.246 & 0.101 & 0.102 & 90 \\
		PIONIER & 57601.215 & 21.69 & 68.08 &  0.42 &  0.33 &   62 \\
		PIONIER & 57900.386 & 22.17 & 68.53 &  0.93 &  0.27 &  144 \\
		PIONIER & 57995.105 & 21.86 & 69.04 &  0.43 &  0.25 &  130 \\
		\hline
	\end{tabular}
\end{table}

\begin{figure}
	\centering
	\includegraphics[width=0.8\columnwidth]{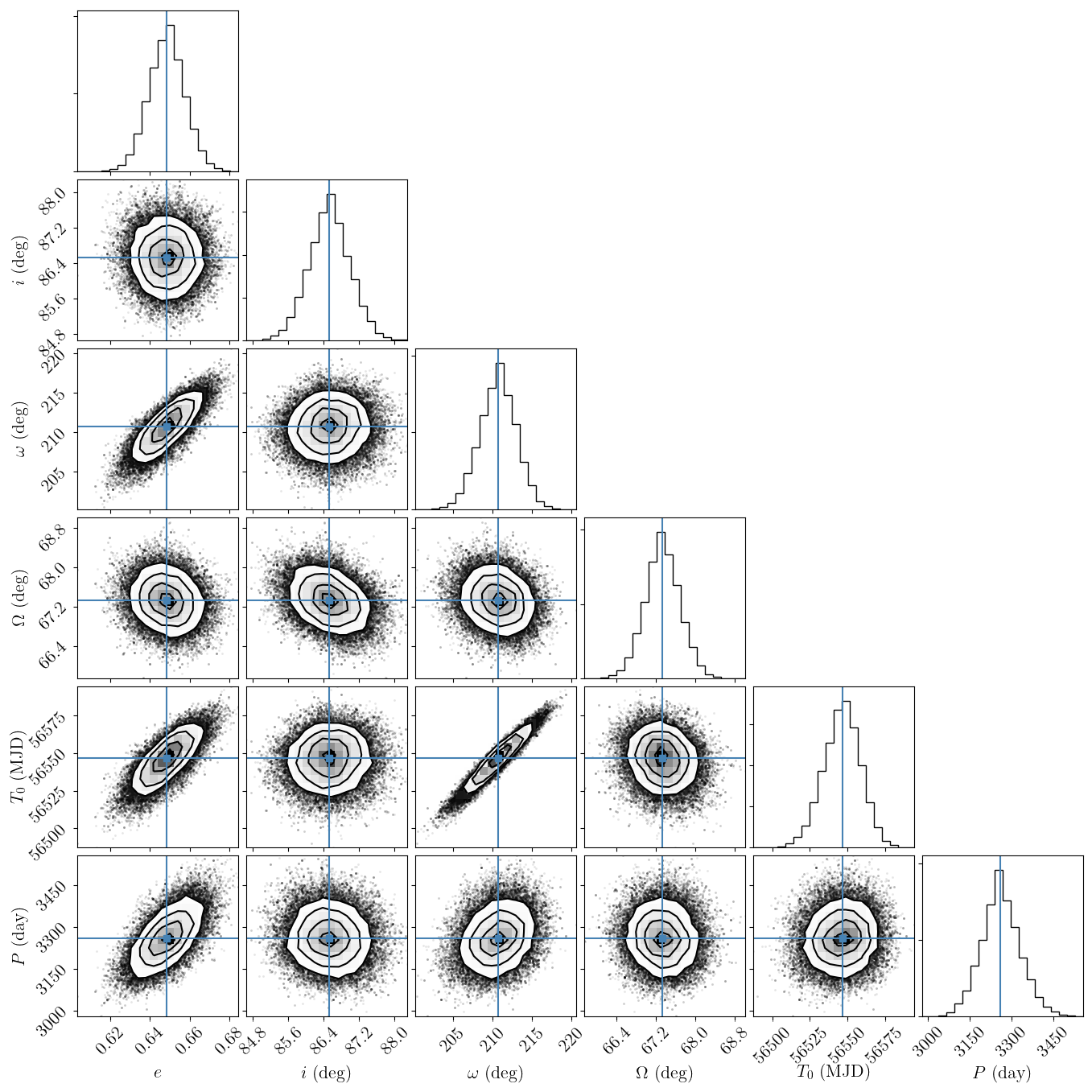}
	\caption{Scatterplot matrix of the MCMC sampling of the orbital solution minimization in section \ref{sec:orbit}.
		While the expected heavy $T_0-\omega$ correlation is apparent, no unexpected degeneracies in the parameter space are observed.}
	\label{fig:corner}
\end{figure}

\begin{figure}
	\centering
	\includegraphics[width=0.5\columnwidth]{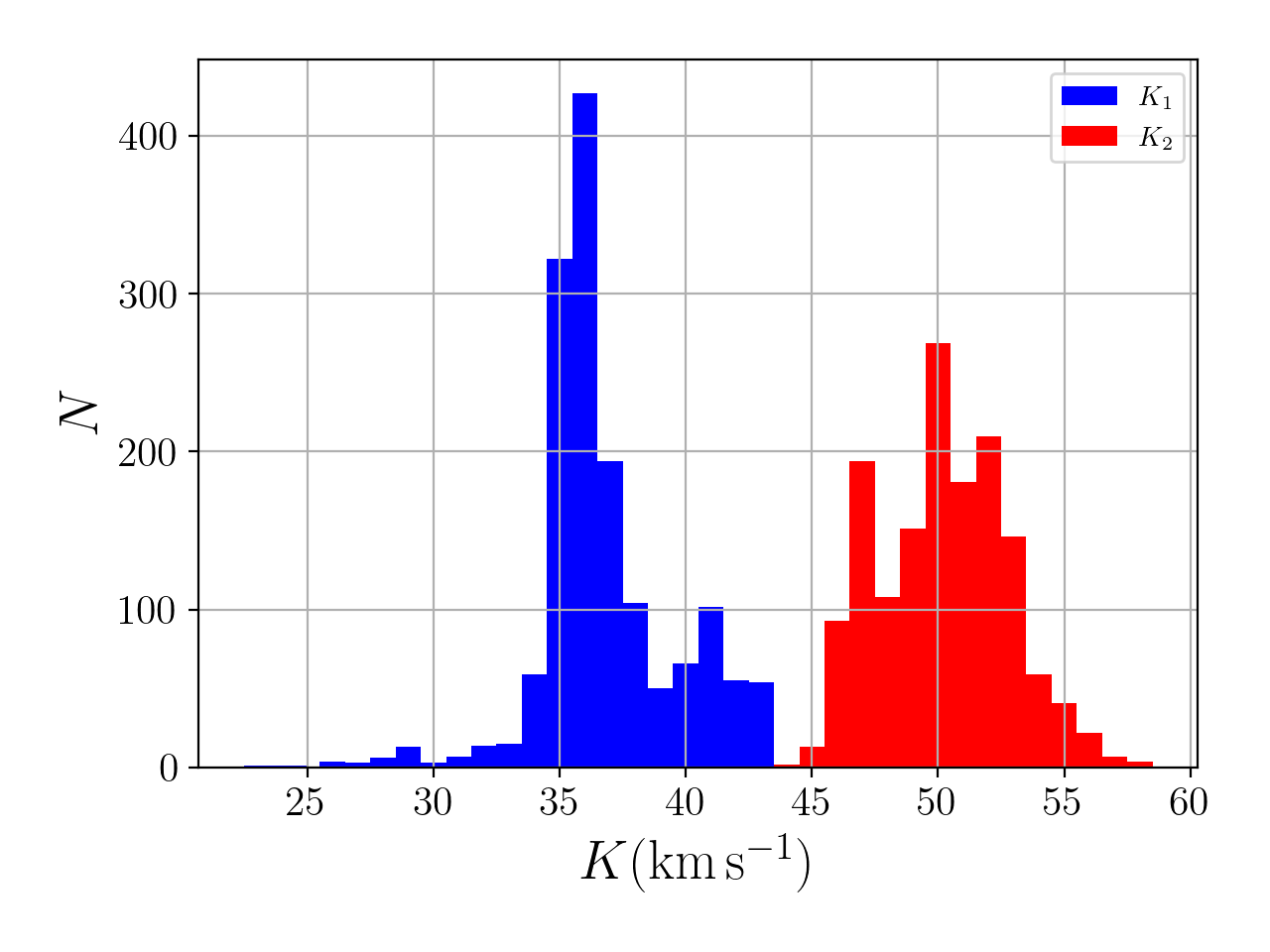}
	\caption{Marginalized histograms of the MC error analysis in Sect.~\ref{sec:disen}.}
	\label{fig:k1k2marg}
\end{figure}

\begin{table}
	\caption{Spectral lines and their respective wavelength ranges, which were fitted against the \textsc{fastwind} atmosphere modeling in Sect.~\ref{sec:atm}, for the primary and secondary star.}
	\label{tab:atmlines}
	\centering
	\begin{tabular}{c c c c}
		\hline\hline
		Spectral line & $\lambda_{\rm min}[\si{\angstrom}]$ & $\lambda_{\rm max}[\si{\angstrom}]$\\
		\hline
		$\ion{He}{i}\,4026$ & 4022.0 & 4029.8 \\
		$\ion{N}{iv}\,4058$ & 4056.5 & 4059.9 \\
		$\element{H}\delta$ & 4093.6 & 4107.9 \\
		$\ion{He}{ii}\,4200$ & 4195.5 & 4204.3 \\
		$\element{H}\gamma$ & 4332.5 & 4348.4 \\
		$\ion{He}{i}\, 4471$ & 4468.6 & 4474.0\\
		$\ion{He}{ii}\, 4541$ & 4535.5 & 4547.3 \\
		$\ion{N}{v}\, 4604$ & 4600.3 & 4608.5 \\
		$\ion{N}{v}\,4619$ & 4617.4 & 4622.8 \\
		$\ion{He}{ii}\, 4686$ & 4680.6 & 4690.4 \\
		$\element{H}\beta$ & 4850.8 & 4870.9\\
		$\ion{He}{ii}\, 5411$ & 5405.4 & 5417.1 \\
		$\ion{O}{iii}\, 5592$ & 5589.3 & 5595.4 \\
		$\ion{C}{iii}\, 5696$ & 5692.4 & 5698.8 \\
		$\ion{C}{iv}\, 5801$ & 5798.8 & 5805.0\\
		$\ion{C}{iV}\, 5812$ & 5809.3 & 5815.0\\
		$\ion{He}{i}\, 5875$ & 5872.0 & 5878.9\\
		\halpha{} & 6551.7 & 6571.2\\
		\hline
	\end{tabular}
\end{table}

\begin{figure}
	\centering
	\begin{minipage}[t]{0.47\columnwidth}
		\centering
		\includegraphics[width=\columnwidth]{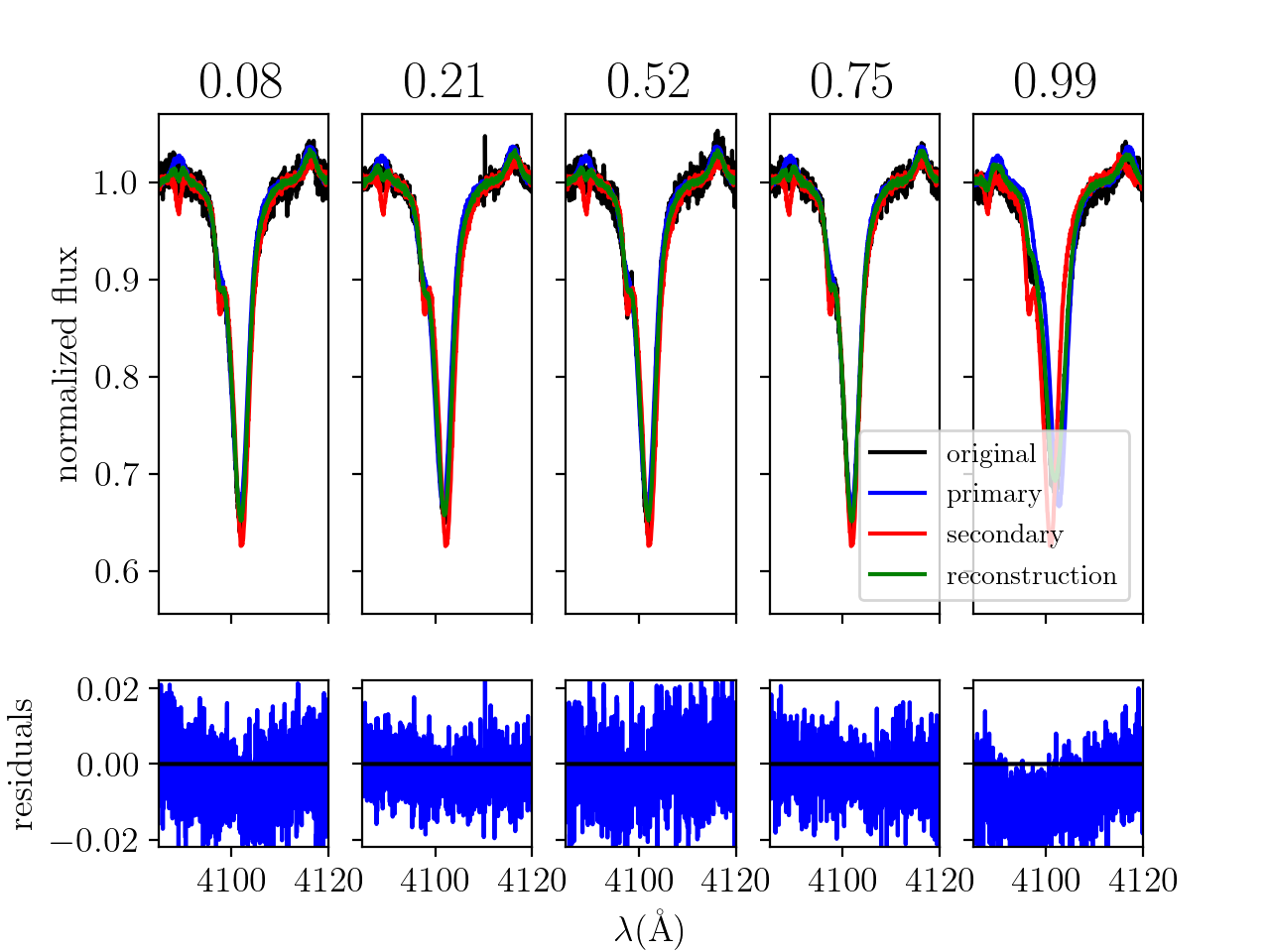}
		\caption{Reconstruction of the $\element{H}\delta$ line at orbital phases indicated above each column.
			At phase 0.99 the reconstruction is slightly too shallow (${\sim}1$\%) near the core, hinting at a possible inconsistent normalization across the spectra.
			The well resolved $\ion{N}{III}$ feature in the left wing of $\element{H}\delta$ is noted.}
		\label{fig:reconhdelta}
	\end{minipage}\hfill
	\begin{minipage}[t]{0.47\columnwidth}
		\centering
		\includegraphics[width=\columnwidth]{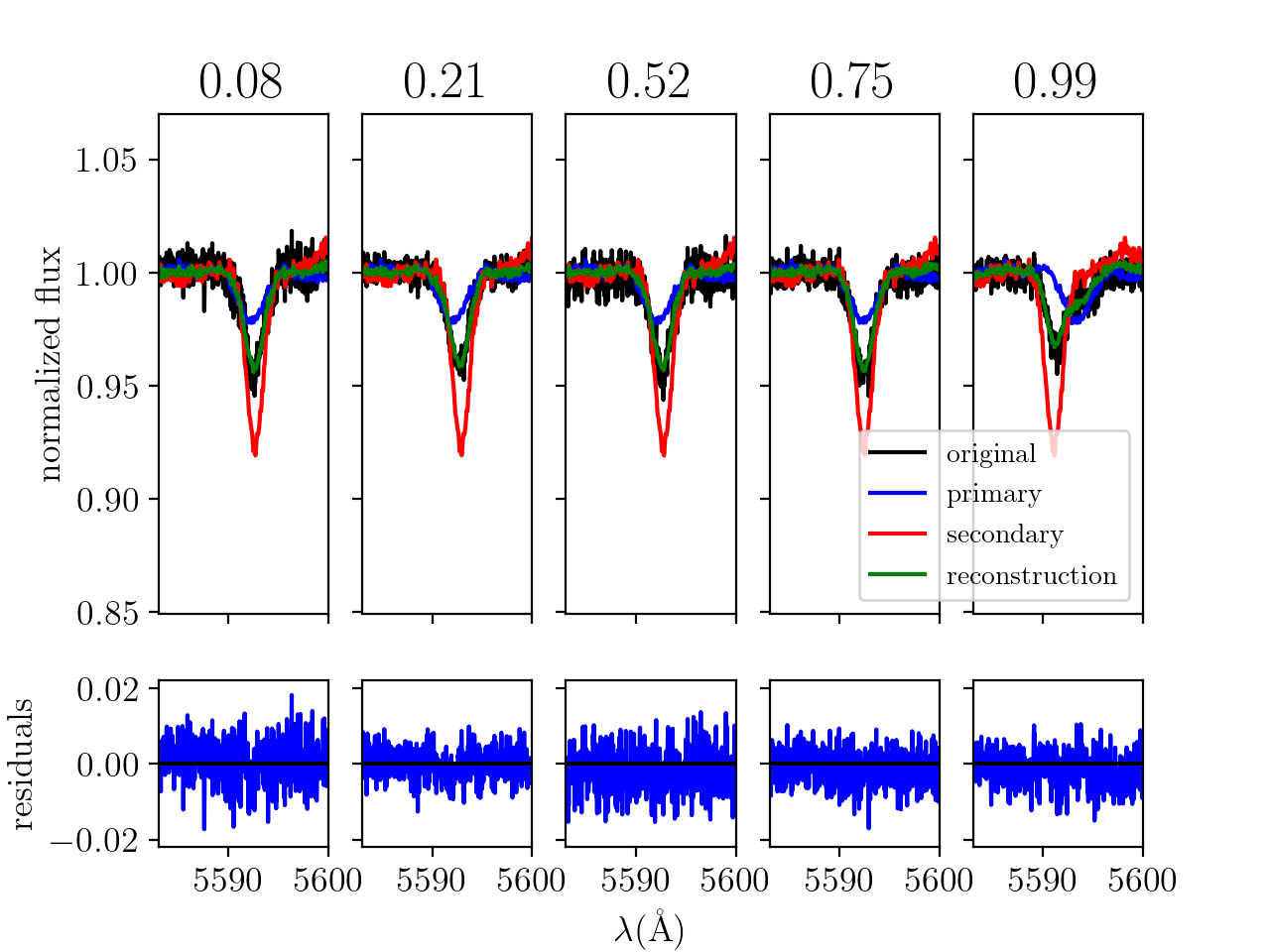}
		\caption{Reconstruction of the $\ion{O}{III}\,5592$ line at orbital phases indicated above each column.
			A good correspondence is observed, even for this relatively weak metal line, although the reconstruction is slightly too shallow at phase 0.99.}
		\label{fig:reconoiii5592}
	\end{minipage}
\end{figure}

\begin{figure}
	\centering
	\includegraphics[width=\columnwidth]{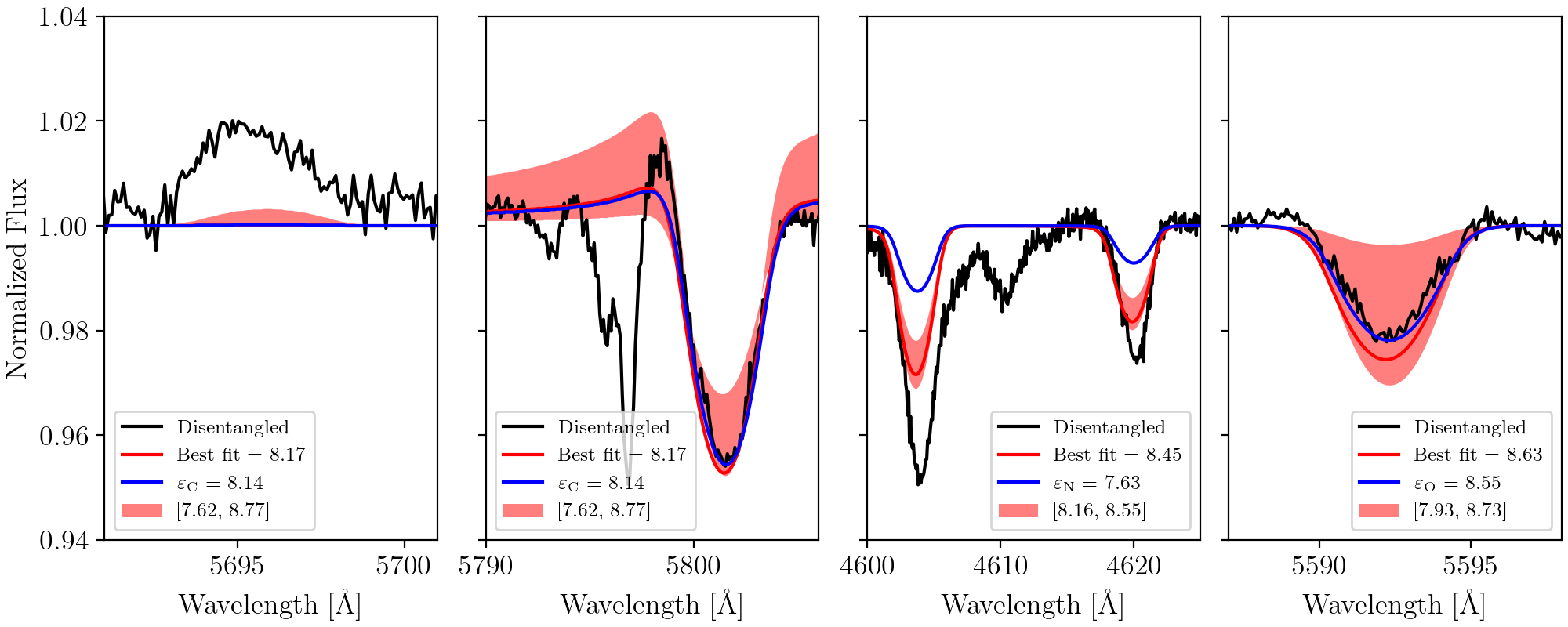}
	\caption{Comparison of the disentangled and best-fit model spectra of the primary star with error ranges, along with a model spectrum having the \citet{brottRotatingMassiveMainsequence2011} CNO baseline abundances.
		From left to right, the lines \ion{C}{iii} $\lambda$ 5696, \ion{C}{iv} $\lambda$ 5801, \ion{N}{v} $\ll{}$ 4604, 4620, and \ion{O}{iii} 5592 are shown, and the legend stipulates the respective CNO abundance values [X/H]+12 and the baseline values $\varepsilon_{\rm X}$.
		As these lines are the main abundance diagnostic for their respective CNO element, from the third panel it can be seen that there is evidence for significant nitrogen enrichment.}
	\label{fig:abunprim}
\end{figure}
\begin{figure}
	\centering
	\includegraphics[width=\columnwidth]{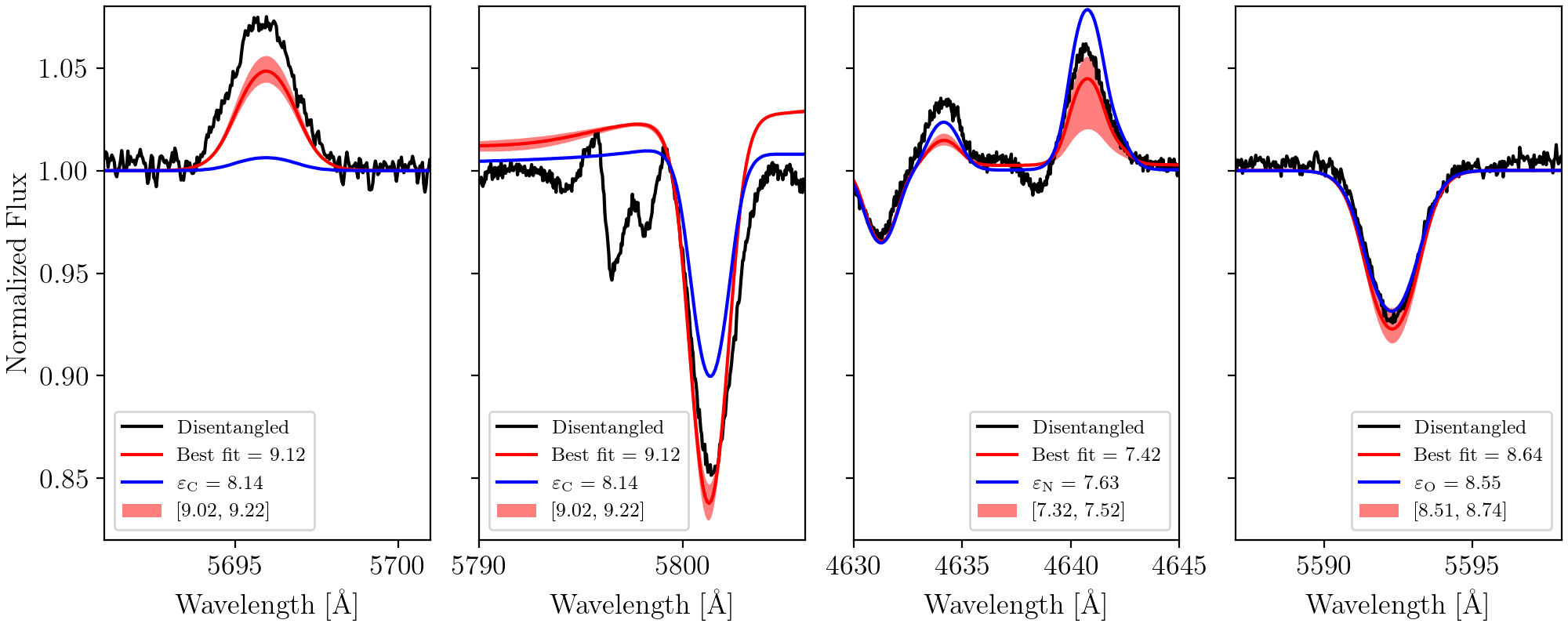}
	\caption{Comparison of the disentangled and best-fit model spectra of the secondary star with error ranges, along with a model spectrum having the \citet{brottRotatingMassiveMainsequence2011} CNO baseline abundances.
		From left to right, the lines \ion{C}{iii} $\lambda$ 5696, \ion{C}{iv} $\lambda$ 5801, \ion{N}{iii} $\ll{}$ 4634, 4641, and \ion{O}{iii} 5592 are shown, with a similar legend as Fig.~\ref{fig:abunprim}.
		The two leftmost panels argue for an elevated carbon abundance with respect to baseline, although as discussed in Sect.~\ref{sec:atmres}, the issues of \ion{C}{iii} $\lambda$ 5696 prohibits making quantitative conclusions. }
	\label{fig:abunsec}
\end{figure}

\FloatBarrier
\twocolumn
\section{{\sc spinOS}}\label{app:spinOS}
In this appendix, we introduce {\sc spinOS}, a modern Python implementation of an orbital minimization algorithm.
The 3D orbit of a binary system can be obtained through two independent sets of measurements: RV measurements via Doppler spectroscopy and apparent separations through interferometric or visual astrometry.\par

We assume the motion of the bodies is governed by Newton's equations and thus Kepler's laws hold ({\it i.e.}, we neglect all relativistic effects).
Both components then follow ellipses described by their radius vector with respect to the common focal point $\vec{r}(t)$, which is a function of time $t$ and is dependent on the orbital elements
\begin{equation}
	\vec{r}(t) = \vec{r}(t; K, e, i, \omega, \Omega, T_0),
\end{equation}
where $K = \frac{2\pi a\sin\,i}{P\sqrt{1-e^2}}$ is the semi-amplitude of the RV curve, $P$ the orbital period, $a$ the semimajor axis, $e$ the eccentricity, $i$ the inclination of the orbital plane with respect to the sky, $\omega$ the argument of periastron, $\Omega$ the longitude of the ascending node, and $T_0$ the time of periastron passage.
In the optimal case in which astrometry and RVs are available, we can determine all the elements as well as the distance to the system.
If we only have astrometry, the (sum of the) semi-amplitudes are degenerate with the distance.
Conversely, if we have only RVs of either or both of the components, the inclination and longitude of the ascending node and the distance remain unconstrained.\par

The RV equation, which measures the velocity of an orbital component along the normal of the plane of the sky ($\equiv z$-axis), is well known and reduces to \citep[see, {\it e.g.},][]{hilditchIntroductionCloseBinary2001}
\begin{equation}
	RV(t) = \dot{z}(t) = K\left(\cos(\theta(t) + \omega) + e\cos\omega\right) + \gamma,
\end{equation}
where $\theta(t)$ is the true anomaly as function of time and $\gamma$ the systemic velocity. The true anomaly relates to time $t$ via the eccentric anomaly $E$ and Kepler's equation as follows:
\begin{align}
	&\tan\left(\frac{\theta}{2}\right) = \sqrt{\frac{1+e}{1-e}}\tan\left(\frac{E}{2}\right),\\
	&E - e\sin E = \frac{2\pi (t-T_0)}{P}.
\end{align}
Computationally, it is the latter of these equations that takes longest to solve because this transcendental equation can only be solved numerically through Newton-Raphson-like iterations or bracketing algorithms.\par
For the astrometric solution, we solve the following equations \citep[see again, {\it e.g.,}][]{hilditchIntroductionCloseBinary2001}, which quantify the relative northward separation $\Delta {\rm N}$ and eastward separation $\Delta {\rm E}$ (in angular units, typically milli-arcseconds) on the plane of the sky:
\begin{align}
	\Delta {\rm N} &= AX + FY,\\
	\Delta {\rm E} &= BX + GY,
\end{align}
where $A, B, F$ and $G$ are the Thiele-Innes constants:
\begin{align}
	A &= a_{\rm app}(\cos\Omega\cos\omega-\sin\Omega\sin\omega\cos i),\label{eq:thielea}\\
	B &= a_{\rm app}(\sin\Omega\cos\omega+\cos\Omega\sin\omega\cos i),\\
	F &= a_{\rm app}(-\cos\Omega\sin\omega-\sin\Omega\cos\omega\cos i),\\
	G &= a_{\rm app}(-\sin\Omega\sin\omega+\cos\Omega\cos\omega\cos i),\label{eq:thieleg}
\end{align}
$a_{\rm app}$ being the semimajor axis of the apparent orbit (in angular units) and $X$ and $Y$ are the rectangular elliptical coordinates defined by
\begin{align}
	X &= \cos E - e,\\
	Y &= \sqrt{1-e^2}\sin E.
\end{align}\par

Given then a set of measurements ${\rm RV1}_{\rm obs}(t_i)$, RV2$_{\rm obs}(t_j)$ and $\Delta {\rm N}_{\rm obs}(t_k)$, $\Delta {\rm E}_{\rm obs}(t_k)$, we minimize the objective function, which doubles as a chi-squared measure of the goodness of fit. Schematically it is given by
\begin{equation}\label{eq:minimization}
	\chi^2(\vartheta) = \sum_i \left(\frac{O(t_i) - C(t_i; \vartheta)}{\sigma(t_i)}\right)^2,
\end{equation}
where $O$ and $C$ are observed and computed data, $\sigma$ the error on the observations, and $\vartheta$ symbolizes the set of orbital parameters.
We use the Levenberg-Marquardt minimization algorithm implemented by the package \texttt{lmfit} \citep{mattnewvilleLmfitLmfitpy2020} to find the orbital parameters $\vartheta$ that minimize eq. \eqref{eq:minimization}.
By default, errors are estimated from the diagonal elements of the covariance matrix, but, an MCMC option is available that takes correlations between parameters into account.
In the latter, a specified number of samples is drawn around the minimum to obtain an approximated posterior probability distribution of the parameters, of which different quantile masses define different confidence intervals.\par

Finally, we present the orbital modeling and minimization routine in a user friendly graphical user interface constructed using \texttt{tkinter}, where the user chooses, among others, which data to be plotted, which orbital parameters to fit for, and which weights to give the astrometry relative to the spectroscopic measurements. The \textsc{spinOS} tool is available under a GNU GPL 3.0 (or later) license through GitHub on \url{https://github.com/matthiasfabry/spinOS}.\par

\end{document}